# Energy Efficient Processing Allocation in Opportunistic Cloud-Fog-Vehicular Edge Cloud Architectures

*Amal A. Alahmadi, T. E. H. El-Gorashi, and Jaafar M. H. Elmirghani*

*Abstract*—This paper investigates distributed processing in Vehicular Edge Cloud (VECs), where a group of vehicles in a car park, at a charging station or at a road traffic intersection, cluster and form a temporary vehicular cloud by combining their computational resources in the cluster. We investigated the problem of energy efficient processing task allocation in VEC by developing a Mixed Integer Linear Programming (MILP) model to minimize power consumption by optimizing the allocation of different processing tasks to the available network resources, cloud resources, fog resources and vehicular processing nodes resources. Three dimensions of processing allocation were investigated. The first dimension compared centralized processing (in the central cloud) to distributed processing (in the multi-layer fog nodes). The second dimension introduced opportunistic processing in the vehicular nodes with low and high vehicular node density. The third dimension considered non-splittable tasks (single allocation) versus splittable tasks (distributed allocation), representing real-time versus non real-time applications respectively. The results revealed that a power savings up to 70% can be achieved by allocating processing to the vehicles. However, many factors have an impact on the power saving such the vehicle processing capacities, vehicles density, workload size, and the number of generated tasks. It was observed that the power saving is improved by exploiting the flexibility offered by task splitting among the available vehicles.

## I. INTRODUCTION

The Cisco Visual Networking Index of 2019 reports that more than six billion M2M (Machine-to-Machine) connections were added in 2017, 28% of these connections are connected vehicles. This number is expected to increase by more than 50% by 2022. This expansion reveals exponential growth in global traffic estimated to exceeds 25 exabyte per month [1]. This growth is accompanied by a remarkable increase in energy consumption in the Information and Communication Technologies (ICT) sector. It is estimated that ICT technologies will be responsible for up to 12% of the global emissions by 2030 [2]. The proliferation of connected devices will lead to rapid growth in the generated traffic between the edge layer and data centers, and therefore is expected to lead to significant increase in the power consumption of the network infrastructure. This calls for new architectural designs capable of reducing the traffic congestion and power consumption in the network. At the same time, vehicles are going through a huge revolution in terms of their on-board units and processing capabilities producing a new promising framework concept. This concept is a consequence of the integration of vehicles and cloud computing, referred to as Vehicular Edge Cloud (VEC). The huge growth in increasingly connected edge devices and resource hungry applications calls for more proposals in distributed processing systems to offload the computational burden in the centralized data centres, and to improve the performance of applications. Distributed cloud platforms can be composed using any available user-owned resources that allow processing, storage, networking and sensing. Following this concept, vehicular clouds can be formed if the vehicle on-board processing, storage, and sensing devices are clustered together to form short-term cloud units composed of many vehicles (each vehicle effectively acting as a server in a mobile micro data centre) at the edge of the network [3]. Furthermore, considering these edge entities as Computing as a Service (CompaaS) providers can help in turning these vehicles from service consumer to cloud-based providers for many applications that are generated from the surrounding connected entities.

In this paper we propose a vehicular edge cloud (VEC) based architecture, where a group of vehicles in a car park, at a charging station or at a traffic signals intersection, cluster and form a temporal vehicular cloud by combining their computational resources in the cluster, as illustrated in Figure 1. The aim of the VEC is to exploit the underutilized computational resources in vehicles' on-board units (OBUs) and to increase the distributed processing resources to serve the demands required by a smart city environment. Vehicle capabilities are expected to grow significantly with the introduction of autonomous vehicles in the near future. Currently, the enterprise parking lot may contain hundreds to thousands of vehicles that remain in the park for typically 7–8 hours per day [3]. If vehicles are connected in such a car park, using wireless connections or a fiber cable integrated with the charging cable and its plug, their processors (typically 2–10 processors per vehicle) can be networked, thus transforming the car park into a significant edge processing micro data centre. The vehicles may alternatively be equipped with a "processing box" that has processing, storage and wireless communication (WiFi for example) capabilities. Such a processing box can reduce the security risks and eliminate the need to connect to the processors in the vehicle or can supplement the vehicle on-board processing capabilities. A set of VECs made up of the parking rows and floors in a car park can thus be formed. Similarly, cars in airports may be parked for one to two weeks, making the capabilities of such vehicles available to transform such car parks to processing units at the edge of the network on a semi-permanent basis as departing cars are replaced. On the shortest time scales, clusters of vehicles may be formed at traffic intersection points where the traffic light may own a computational problem and may assign chunks of such a computational problem to vehicle clusters at the intersection. The clusters report results before departing the intersection. At busy intersections in cities, typically at least one traffic stream is stationary, thus providing opportunities to distribute computational tasks to nearby processors. These vehicles thus have the potential to form efficient short-term distributed computational resources at the edge of the network, much closer to the requesting entity. With vehicle availabilities that can range from minutes to weeks, the networking and computational resources are highly dynamic. Therefore, appropriate network architectures and network algorithms are needed to better utilize these new forms of dynamic distributed computational resources.

The proposed architecture, in Figure 1, is integrated into a multi-layer fog mini data centre and supported by a central cloud data centre. All the considered cloud-fog-edge resources, referred to as processing nodes (PNs), act as a service provider for smart city demands for example. These processing demands are assumed to be generated from IoT sensor nodes distributed across the streets and in the city near car parks, charging stations or road intersections. The main goal of the proposed framework is to find the optimum processing task placement (to allocate and process the generated tasks) in order to minimize the total power consumption of the end-to-end architecture. This work studies the joint energy efficiency of network and processing along three dimensions. The first dimension compares the centralized processing (in the central cloud) to distributed processing (in the multi-layer fog nodes). The second dimension introduces opportunistic processing in the vehicular nodes. The third dimension considers non-splittable tasks (single allocation) versus splittable tasks (distributed allocation), representing real-time versus non real-time applications.

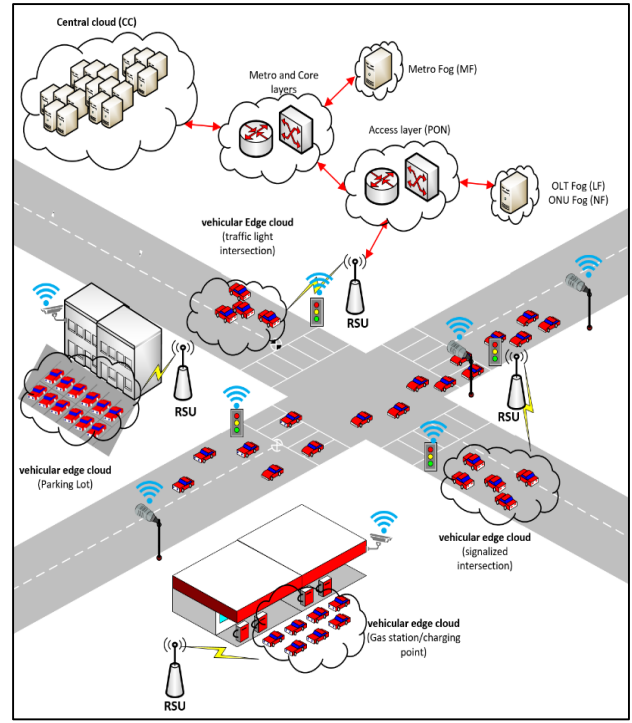

**Figure 1** Cloud-Fog-Vehicular Edge Cloud Architecture

## II. RELATED WORK

The main clouds in the Internet use thousands of servers and hardware from several data centres to process all the application requests that come from users. Cloud applications that are hosted by these data centres consume a large amount of energy for processing and for cooling the hardware [4]. Power consumption has been escalating along with the rapidly increasing use of cloud infrastructure. It is estimated that ICT technologies will be responsible for up to 12% of global emissions by 2030 [2]. Accordingly, issues relating to greening the ICT sector have received more attention in recent years. There is a growing recognition of the need for more research to develop green cloud computing infrastructure in order to save energy and reduce the negative impact of this power expenditure on the environment. Providing energy efficient and reliable network infrastructure has been the main focus of attention in recent years, which has highlighted the need for more research to develop new architecture and solutions that will provide robust and energy efficient infrastructure. Previous research efforts have contributed solutions that can be used to reduce the power consumption of cloud data centres and core networks [5]–[9]. Different techniques and technologies are considered to improve network energy efficiency including virtualization [10]–[12], network architecture design and optimization [13]–[16], optimized content distribution [17]–[19], big data progressive processing and edge processing [20]–[23], network coding to reduce the volume of data in the network [24], [25] and renewable energy to reduce the carbon footprint of the network[26].

Decentralized architectures have also been proposed to integrate distributed edge servers to mitigate the traffic burden on central data centers. Recent proposed solutions include building decentralized architectures that integrate distributed edge servers in order to mitigate the traffic burden on central data centres, and thus save more energy [27]. Moreover, as these distributed data centres provide access to the computational resources in distributed servers at the edge of the network, they provide cloud-based services that are in close proximity to the end user. Hence, edge computing offers a good solution for the conventional cloud to offload its processing workload to these distributed servers and therefore, save power. Thus, it is very important to make sure that as these mini data centres are built, it is taken into consideration that they should not significantly increase the power overhead in the ICT sector. Based on this idea, many research efforts have been focused on investigating different architectural and network designs. Moreover, efforts have been made to investigate and solve optimization problems in resource management and workload allocation in order to achieve energy efficient data centres that are at the edge of the network. Most of the studies already conducted have focused on fixed distributed servers. For example, in the study in [28], the authors studied and analyzed the energy efficiency of processing applications in nano data centres compared to those in the central cloud. Their study shows that there are many factors that affect the efficiency of the nano data centre, such as the server's location, the equipment in the access network, and the number of user requests. Other studies, however, have focused on utilizing available ICT resources, such as nano datacenters, considering the power consumed in such a framework. This research area mainly focuses on utilizing distributed resources, such as IoT nodes [29], smartphones [30], and other mobile devices, including vehicles. Many researchers have also focused on the area of mobile cloud computing to investigate energy consumption. The main concerns regarding this framework involve the limited power resources of the mobile devices and the ways in which they can be used as distributed processing units, given their limited power supply [30]. Other efforts relating to energy consumption in mobile cloud computing (MCC) are summarized in [31]. Fewer studies have focused on investigating energy efficient network proposals, including those using vehicles as distributed servers. Our previous efforts in the area of vehicular cloud have started with a preliminary optimization model to investigate the power consumed in central cloud compared to the vehicular cloud [32]. This work was extended to study the optimization problem considering multiple allocation strategies [33], software matching [34], [35], delay [36], [37], and an optimized vision to incorporate vehicular cloud in future networks [38].

Most of the other optimization efforts in the vehicular cloud focus on providing a distributed system with efficient, cost-accurate models by either minimizing the operational cost or maximizing the reward cost for the owners of the servers. As rewards and the associated cost structures are beyond the scope of this paper, we will summarize some of the optimization models that involve a proposed cost-minimization model. In addition, as energy is considered one of the cost metrics, we will discuss the work that focused on minimizing the power consumption of networks in vehicular clouds (VC). Moreover, we will highlight the few studies that investigate energy as an optimization measure in VC. Then, we will give a brief summary of the work that focuses on maximizing user rental costs.

The work in [39] proposed a generic optimization model for storage allocation in parked vehicles to minimize the total communication costs. The allocation problem considered forwarding video-on-demand requests to vehicles that had the required video available. The minimized communication costs included request service costs and management costs. The video requested was allocated to a vehicle based on the probability of that vehicle having the requested video cache available. The authors' proposed optimization model was mathematically analyzed and developed through multiple management policies based on replicating (or not replicating) the video-cached copies among the available vehicles. Their results conclude that the communication costs are reduced by increasing the number of available vehicles, thereby reducing the number of downloaded cache replicas. This work only considered the costs of downloading the video cache into the vehicle. Moreover, no further details were given about the optimization model, the developed algorithm, or the system's infrastructure.

In a similar work, the authors of [40] optimized job allocation based on the requested service type. The developed model minimizes the job completion cost, which is a function of the communication and computing costs of the allocated task. The total cost was formulated using mixed integer non-linear programming (MINLP). The proposed allocation relies on choosing the optimum vehicle to offload the processing workload from the road-side unit (RSU), considering pre-defined predicted vehicle mobility algorithms. The proposed optimization model was proven to have better total cost compared to a conventional case, where each processing job was allocated to all available vehicles surrounding the RSU. This work found that in-vehicle processing reduces the burden of the RSU. However, the proposed system cannot guarantee reliable service delivery by considering processing only from opportunistic vehicles or the limited-resourced RSU.

The authors of [41] also proposed a generic cost minimization based on a time-scheduling optimization model in order to satisfy task completion. They executed the model using binary integer programming (BIP) and included both processing and networking costs, taking into account two available communication mediums (WAVE and 3G / 4G). They consider tasks that are executed in independent and parallel rounds in order to capture different time slots and to study dynamic resource variability (i.e. vehicles' resources). The model works under the assumption that each task is processed only in one location, which restricts the type of application that can be accommodated in such a system. The findings of the work show that vehicles with a WAVE connection can offer low-cost processing. However, connection stability and limited capacity can be an issue in terms of service reliability.

Few studies have investigated energy efficiency in VC-based optimization models. However, in [42], the authors tackled the problem of offloading from smartphones to another processing node in order to reduce the computational overhead of the limited smartphone processors. They developed a flexible offloading algorithm to optimize the task-computing placement between the smartphone (locally), central cloud, cloudlet server, or VC. In each allocation placement, they assessed the energy consumed and response time with multiple capacities of processing locations and multiple input sizes. Their proposed algorithm gives priority to processing tasks in the central cloud and in cloudlet locations. It offloads the task to VC only in cases where both other locations do not satisfy the capacity or time demands of the task. Their findings show that the allocation of tasks to VC creates the lowest delay in most input sets (except for those with very high-demand traffic) and achieves the lowest response time compared to the other processing locations.

The authors in [43] presented another attempt to minimize energy consumption by optimizing the task-offloading decision in the VC framework. Their proposed framework consists of distributed servers in RSU and collocated vehicles, with each RSU also working as a processing resource. The vehicles are assumed to be a user that generates requests to be processed locally or offloaded to an RSU or another vehicle if the request exceeds its limited processor capacity. The model optimizes this decision by minimizing the total energy consumed, which includes the local processing energy, the offload transmission energy, and the processing energy in either the RSU or the vehicle's processor. In addition, the model considers the processing time constraints of some of the generated tasks. An extensive mathematical analysis was given of the offloading problem, and energy consumption was analyzed based on the portion of the offloaded workload and the assumed transmission power of the RSU and vehicles' access point.

Using a different approach, some researchers developed a joint optimization model that has more than one objective. For example, the authors in [44] proposed a fitness estimation model to allocate tasks to a network of three layers, which include the central cloud, cloudlet nodes, and VC. This model was designed using three objectives to minimize network delay, minimize power consumption, and maximize the availability of the processing resources (as virtual machines). The total network delay is calculated considering the propagation, queuing, transmission, and processing delays. The queuing delay considered is based on the waiting time of the processing tasks in each processor and is calculated using the total execution time. This is the same for the power consumption, which is calculated only based on the power needed to execute each allocated task. The study mainly focused on the proposed fitness estimation model, which studied the capability of the available resources and their availability, a problem that is considered a challenge due to the vehicles' mobility.

The remainder of this paper is organized as follows: In Section III, we introduce the proposed end-to-end cloud-fog-VEC architecture. In Section IV, the optimization model is described. The input and parameters are described in Section V. In Section VI, the scenarios considered and the results are discussed. Finally, Section V concludes the paper and highlights some of the planned future work.

## III. THE PROPOSED CLOUD-FOG-VEHICULAR EDGE CLOUD ARCHITECTURE

The proposed integrated cloud-fog-VEC end-to-end architecture is shown in Figure 2. It is composed of four distinct layers with four processing locations at core, metro, access, and edge layers.

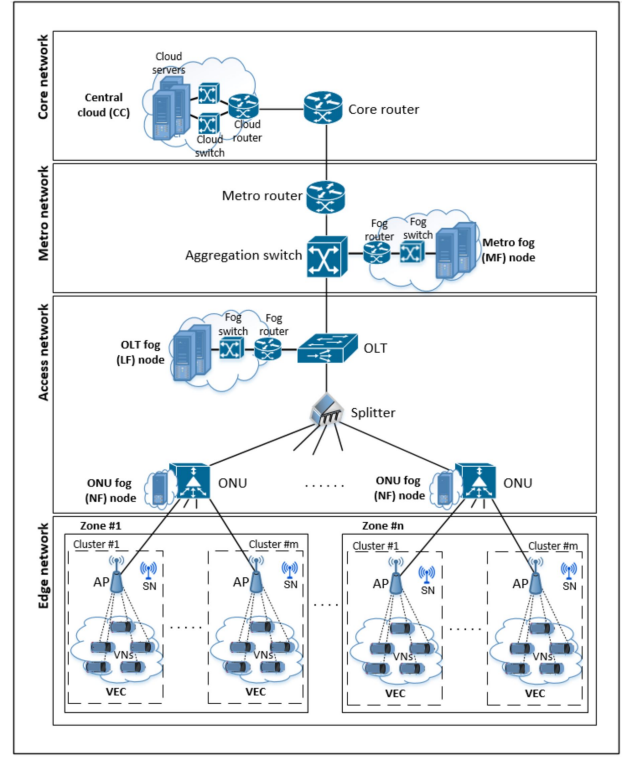

**Figure 2** End-to-End Cloud-Fog-VEC Architecture

### A. Edge network

This network forms the bottom-most layer and represents a snapshot of a smart city layout. Two types of edge entities are defined in this layer, namely source nodes (SNs) and vehicular nodes (VNs). SNs are distributed IoT nodes (for example wireless sensors) which are usually responsible for collecting multimedia and environmental data to generate information used by security and environmental monitoring applications. The edge network also includes one or more temporal VNs clustered in car parks, charging stations or road intersections forming a VEC. These vehicles are equipped with OBUs and can work as a processing node to process and analyze the collected data. Both IoT SNs and VNs are connected to the wired infrastructure through an Access Point (AP). The AP acts as a controller that collects the IoT-generated tasks and allocates them to the optimum PNs. The communication medium between the AP and edge nodes (IoTs and VNs) is selected according to the controller unit (AP) and the communication protocol in the edge nodes (i.e. wireless connection). This AP is assumed to have full knowledge of the available resources and has enough computational capability to fulfil its coordination and allocation roles.

As seen in Figure 2, the design of the edge layer is based on multiple zones, where each zone represents one geographical area. Each zone may also include one or more VEC. Every VEC is represented by VNs clustered in a car park, at a charging station or at an intersection and an AP. VNs within the same VEC can communicate only with one local AP. As the AP has the role of collecting and allocating tasks, it can communicate with other VEC clusters through the access network via an Optical Network Unit (ONU). Moreover, tasks generated from one zone can also be allocated to other zones, through a Passive Optical Network (PON) and via an Optical Line Terminal (OLT). The PON design, including access layer entities, will be explained next.

### B. Access network

This layer consists of a PON with several ONU devices, each connected to the AP devices distributed in the same zone. These ONUs are connected to an OLT via a fibre link using a passive optical splitter. Fixed fog processing nodes can be deployed at both ONUs and the OLT, named ONU fog (NF) and OLT fog (LF), respectively. Processing nodes located at the ONU are small and limited in their processing capability, but provide a closer processing opportunity to the edge source nodes. The former nodes also provide more reliable processing nodes that serve processing demands which cannot be satisfied by the VEC. On the other hand, a processing node located at the OLT has more processing capabilities compared to the NF and VEC, and is considered as a supportive processing layer for the generated demands to guarantee reliable service provision.

### C. Metro network

The metro layer is an intermediate network between the access layer and the core layer. It consists of a switch, which aggregates the data collected from the edge and access layers, and a metro router, while simultaneously serving as a gateway between the access layer and core layer. Another fixed fog is included in this layer and is connected to the aggregation switch. This metro fog (MF) node is equipped with servers that have higher computational capabilities compared to the previously explained processing nodes

### D. Core network

This layer includes core routers connected to the central cloud that has its own cloud routers, cloud switches, and cloud servers. The central cloud supports the architecture with servers that have high processing capabilities to execute tasks that cannot be executed in the lower processing nodes.

It is worth mentioning that all four layers are scalable. For model complexity runtime, we assumed one PN in each cloud and fog layer, except NF (in some scenarios) and VEC (as we are assessing the opportunistic behavior of the VNs through VNs density).

## IV. MILP MODEL

This section introduces the Mixed Integer Linear Programming (MILP) model that has been developed to minimize the total power consumption by optimizing the processing allocation of different demands into the available processing locations in the integrated cloud-fog-VEC architecture.

As the total power consumption includes all devices involved in processing the demands or networking the associated traffic, it is necessary to describe the power profile of these devices and how it is related to the generated processing/traffic demands. The power profile considered in this work is based on a linear power profile [45]. Hence, the power consumption of all network equipment, including processing nodes, consists of a linear proportional part and an idle part, as shown in Figure 3. It is very important to consider the idle power consumption ($P_{idle}$), as it can represent a large percentage of the maximum power consumption ($P_{max}$), typically 60%-95% [45]. Hence, the total power consumption calculated in the model considers these values to calculate the power per processor MIPS and the power per bit/sec, using:

$$P_L = P_{idle} + L \left(\frac{P_{max} - P_{idle}}{C_{max}}\right) \quad (1)$$

where $C_{max}$ is the maximum workload the device can handle, $L$ is the workload allocated to the processing or networking device, $P_{idle}$ is the idle power of the device when, $L = 0$, and $P_{max}$ is the power consumption when the workload $L = C_{max}$.

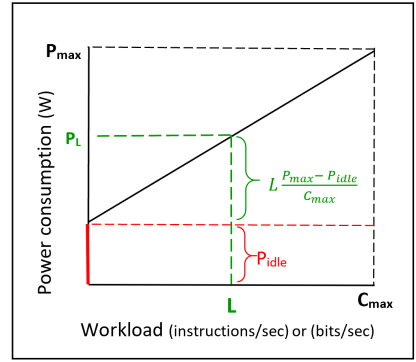

Fig. 3. Power consumption profile

The sets, parameters, and variables are declared as follows:

| Set: | |
|---|---|
| $N$ | Set of all nodes. |
| $NN_i$ | Set of neighbors of node $i$, $\forall\ i \in N$. |
| $PN$ | Set of processing nodes, where $PN \subset N$. |
| $SN$ | Set of source nodes, where $SN \subset N$. |
| $RR$ | Set of core router ports, where $RR \subset N$. |
| $MR$ | Set of metro router ports, where $MR \subset N$. |
| $MS$ | Set of metro switches, where $MS \subset N$. |
| $O$ | Set of OLT nodes, where $O \subset N$. |
| $U$ | Set of ONU nodes, where $U \subset N$. |
| $A$ | Set of AP nodes, where $A \subset N$. |
| $CC$ | Set of central cloud (CC) servers, where $CC \subset PN$. |
| $CR$ | Set of CC router ports, where $CR \subset N$. |
| $CS$ | Set of CC switches, where $CS \subset N$. |
| $MF$ | Set of Metro fog servers, where $MF \subset PN$. |
| $MFR$ | Set of MF router ports, where $MFR \subset N$. |
| $MFS$ | Set of MF switches, where $MFS \subset N$. |
| $LF$ | Set of OLT fog servers, where $LF \subset PN$. |
| $LFR$ | Set of LF router ports, where $LFR \subset N$. |

| Symbol | Description |
|---|---|
| $LFS$ | Set of LF switches, where $LFS \subset N$. |
| $NF$ | Set of ONU fog processors, where $NF \subset PN$. |
| $VN$ | Set of vehicular nodes processors, where $VN \subset PN$. |

**Parameters:**

| Symbol | Description |
|---|---|
| $\omega_s$ | Processing requirement of the task generated from source node $s \in SN$, in Million Instructions per Seconds (MIPS). |
| $\mathcal{F}_s$ | Traffic flow demand generated from source node $s \in SN$ (in Mb/s). |
| $F_s$ | Traffic flow to processing demand ratio for source node $s \in SN$, where $F_s = \frac{\mathcal{F}_s}{\omega_s}$. |
| $C_d$ | Maximum capacity of processing node $d \in PN$ (in MIPS). |
| $L_{ij}$ | Maximum capacity of the link between nodes $(i,j)$ (in Mb/s), where $i \in N$, $j \in NN_i$. |
| $P_{max}^{RR}$ | Core router port maximum power consumption (W). |
| $P_{max}^{MR}$ | Metro router port maximum power consumption (W). |
| $P_{max}^{MS}$ | Metro switch maximum power consumption (W). |
| $P_{max}^{O}$ | OLT maximum power consumption (W). |
| $P_{max}^{U}$ | ONU maximum power consumption (W). |
| $P_{max}^{A}$ | AP maximum power consumption (W). |
| $P_{max}^{CC}$ | CC server maximum power consumption (W). |
| $P_{max}^{CR}$ | CC router port maximum power consumption (W). |
| $P_{max}^{CS}$ | CC switch maximum power consumption (W). |
| $P_{max}^{MF}$ | MF server maximum power consumption (W). |
| $P_{max}^{MFR}$ | MF router port maximum power consumption (W). |
| $P_{max}^{MFS}$ | MF switch maximum power consumption (W). |
| $P_{max}^{LF}$ | LF server maximum power consumption (W). |
| $P_{max}^{LFR}$ | LF router port maximum power consumption (W). |
| $P_{max}^{LFS}$ | LF switch maximum power consumption (W). |
| $P_{max}^{NF}$ | NF processor maximum power consumption (W). |
| $P_{max}^{VN}$ | VN processor maximum power consumption (W). |
| $P_{max}^{VW}$ | VN wireless adapter maximum power consumption (W). |
| $P_{idle}^{RR}$ | Core router port idle power consumption (W). |
| $P_{idle}^{MR}$ | Metro router port idle power consumption (W). |
| $P_{idle}^{MS}$ | Metro switch idle power consumption (W). |
| $P_{idle}^{O}$ | OLT idle power consumption (W). |
| $P_{idle}^{U}$ | ONU idle power consumption (W). |
| $P_{idle}^{A}$ | AP idle power consumption (W). |
| $P_{idle}^{CC}$ | CC server idle power consumption (W). |
| $P_{idle}^{CR}$ | CC router port idle power consumption (W). |
| $P_{idle}^{CS}$ | CC switch idle power consumption (W). |
| $P_{idle}^{MF}$ | MF server idle power consumption (W). |
| $P_{idle}^{MFR}$ | MF router port idle power consumption (W). |
| $P_{idle}^{MFS}$ | MF switch idle power consumption (W). |
| $P_{idle}^{LF}$ | LF server idle power consumption (W). |
| $P_{idle}^{LFR}$ | LF router port idle power consumption (W). |
| $P_{idle}^{LFS}$ | LF switch idle power consumption (W). |
| $P_{idle}^{NF}$ | NF processor idle power consumption (W). |
| $P_{idle}^{VN}$ | VN processor idle power consumption (W). |
| $P_{idle}^{VW}$ | VN wireless adapter idle power consumption (W). |
| $\Omega_{max}^{RR}$ | Core router port maximum capacity (Mb/s). |
| $\Omega_{max}^{MR}$ | Metro router port maximum capacity (Mb/s). |
| $\Omega_{max}^{MS}$ | Metro switch maximum capacity (Mb/s). |
| $\Omega_{max}^{O}$ | OLT maximum capacity (Mb/s). |
| $\Omega_{max}^{U}$ | ONU maximum capacity (Mb/s). |
| $\Omega_{max}^{A}$ | AP maximum capacity (Mb/s). |
| $\Omega_{max}^{CC}$ | CC server maximum processing capacity (MIPS). |
| $\Omega_{max}^{CR}$ | CC router port maximum capacity (Mb/s). |
| $\Omega_{max}^{CS}$ | CC switch maximum capacity (Mb/s). |
| $\Omega_{max}^{MF}$ | MF server maximum processing capacity (MIPS). |
| $\Omega_{max}^{MFR}$ | MF router port maximum capacity (Mb/s). |
| $\Omega_{max}^{MFS}$ | MF switch maximum capacity (Mb/s). |
| $\Omega_{max}^{LF}$ | LF server maximum processing capacity (MIPS). |
| $\Omega_{max}^{LFR}$ | LF router port maximum capacity (Mb/s). |
| $\Omega_{max}^{LFS}$ | LF switch maximum capacity (Mb/s). |
| $\Omega_{max}^{NF}$ | NF processor maximum processing capacity (MIPS). |
| $\Omega_{max}^{VN}$ | VN processor maximum processing capacity (MIPS). |
| $\Omega_{max}^{VW}$ | VN wireless adapter maximum capacity (Mb/s). |
| $\rho^{CC}$ | Central cloud PUE. |
| $\rho^{MF}$ | Metro fog node PUE. |
| $\rho^{LF}$ | OLT fog node PUE. |
| $\rho^{NET}$ | Network devices PUE. |
| $\tau$ | Portion of the network devices idle power attributed to the application. |
| $\upsilon$ | Number of splits a task can be divided into. |

**Variables:**

| Symbol | Description |
|---|---|
| $X_{sd}$ | Processing workload, in MIPS, generated from source node $s \in SN$ and allocated to processing node $d \in PN$. |
| $\delta_{sd}$ | Binary variable, $\delta_{sd} = 1$ if workload generated from source node $s \in SN$, is allocated to processing node $d \in PN$, 0 otherwise. |
| $\delta_d$ | Binary variable, $\delta_d = 1$ if any workload is allocated to processing node $d \in PN$, 0 otherwise. |
| $\lambda_{sd}$ | Traffic flow sent from source node $s \in SN$ to processing node $d \in PN$. |
| $\lambda_j$ | Total traffic in node $j \in N$ |
| $\lambda_{ij}^{sd}$ | Traffic flow sent from source node $s \in SN$ to processing node $d \in PN$ through physical link nodes $(i,j)$ (in Mb/s), where $s \in SN$, $d \in PN$, $i \in N$, $j \in NN_i$. |
| $\Psi_i$ | Binary variable, $\beta_i = 1$ if any traffic traverses network node $i \in N$, 0 otherwise. |
| $M1$ | Large enough number with unit of MIPS. |
| $M2$ | Large enough number with unit of MIPS. |
| $M3$ | Large enough number with unit of Mb/s. |
| $TPC^{CC}$ | Total power consumption of CC. |
| $PPC^{CC}$ | Processing power consumption of CC. |
| $NPC^{CC}$ | Networking power consumption of CC. |
| $TPC^{MF}$ | Total power consumption of MF. |
| $PPC^{MF}$ | Processing power consumption of MF. |
| $NPC^{MF}$ | Networking power consumption of MF. |
| $TPC^{LF}$ | Total power consumption of LF. |
| $PPC^{LF}$ | Processing power consumption of LF. |
| $NPC^{LF}$ | Networking power consumption of LF. |
| $TPC^{NF}$ | Total power consumption of NF. |
| $TPC^{VN}$ | Total power consumption of VN. |
| $PPC^{VN}$ | Processing power consumption of VN processor. |
| $NPC^{VN}$ | Networking power consumption of VN wireless adapter. |
| $TPC^{NET}$ | Total power consumption of the infrastructure network. |
| $TPC^{RR}$ | Total power consumption of core router. |
| $TPC^{MR}$ | Total power consumption of metro router. |
| $TPC^{MS}$ | Total power consumption of metro switch. |
| $TPC^{O}$ | Total power consumption of OLT. |
| $TPC^{U}$ | Total power consumption of ONU. |

| $TPC^A$ | Total power consumption of AP. |

The total power consumption is composed of the following:

**1) The total power consumption of CC** ($\text{TPC}^{CC}$), which is composed of the processing power consumption ($PPC^{CC}$) and the networking power consumption ($NPC^{CC}$), and given as:

$$\text{TPC}^{CC} = (PPC^{CC} + NPC^{CC})\, \rho^{CC} \quad (2)$$

where $\rho^{CC}$ is the PUE of the central cloud data centre.

$$PPC^{CC} = \left[ P_{idle}^{CC} \sum_{d \in CC} \delta_d + \frac{P_{max}^{CC} - P_{idle}^{CC}}{\Omega_{max}^{CC}} \left( \sum_{s \in SN} \sum_{d \in CC} X_{sd} \right) \right] \quad (3)$$

$$NPC^{CC} = \left[ \tau P_{idle}^{CR} \sum_{i \in MCR} \Psi_i + \left(\frac{P_{max}^{CR} - P_{idle}^{CR}}{\Omega_{max}^{CR}}\right) \left(\sum_{i \in CR} \lambda_i\right) \right]$$
$$+ \left[ \tau P_{idle}^{CS} \sum_{i \in CS} \Psi_i + \left(\frac{P_{max}^{CS} - P_{idle}^{CS}}{\Omega_{max}^{CS}}\right)\left(\sum_{i \in CS} \lambda_i\right) \right] \quad (4)$$

Equation (3) determines the processing power consumption of CC servers. Equation (4) determines the networking power consumption which is composed of the power consumption of the routers (CR) and switches (CS) of the CC network. Note that $\tau$ here represents the portion of the idle power attributed to the considered application traffic, which is equal to 6% [46]. More details will be given later in Section V.

**2) The total power consumption of the MF** ($\text{TPC}^{MF}$), which is composed of the processing power consumption ($PPC^{MF}$) and the networking power consumption ($NPC^{MF}$), and is given as:

$$\text{TPC}^{MF} = (PPC^{MF} + NPC^{MF})\, \rho^{MF} \quad (5)$$

where $\rho^{MF}$ is the PUE of the Metro Fog node.

$$PPC^{MF} = \left[ P_{idle}^{MF} \sum_{d \in MF} \delta_d + \left(\frac{P_{max}^{MF} - P_{idle}^{MF}}{\Omega_{max}^{MF}}\right) \left(\sum_{s \in SN} \sum_{d \in MF} X_{sd}\right) \right] \quad (6)$$

$$NPC^{MF} = \left[ \tau P_{idle}^{MFR} \sum_{i \in MFR} \Psi_i + \left(\frac{P_{max}^{MFR} - P_{idle}^{MFR}}{\Omega_{max}^{MFR}}\right) \left(\sum_{i \in MFR} \lambda_i\right) \right]$$
$$+ \left[ \tau P_{idle}^{MFS} \sum_{i \in MFS} \Psi_i + \left(\frac{P_{max}^{MFS} - P_{idle}^{MFS}}{\Omega_{max}^{MFS}}\right) \left(\sum_{i \in MFS} \lambda_i\right) \right] \quad (7)$$

Equation (6) determines the processing power consumption of the MF server. Equation (7) determines the networking power consumption which is composed of the power consumption of the routers (MFR) and switches (MFS) of the MF network.

**3) The total power consumption of the LF** ($\text{TPC}^{LF}$), which is composed of the processing power consumption ($PPC^{LF}$) and the networking power consumption ($NPC^{LF}$), and is given as:

$$\text{TPC}^{LF} = (PPC^{LF} + NPC^{LF})\, \rho^{LF} \quad (8)$$

$$PPC^{LF} = \left[ P_{idle}^{LF} \sum_{d \in LF} \delta_d + \left(\frac{P_{max}^{LF} - P_{idle}^{LF}}{\Omega_{max}^{LF}}\right) \left(\sum_{s \in SN} \sum_{d \in LF} X_{sd}\right) \right] \quad (9)$$

$$NPC^{MF} = \left[ \tau P_{idle}^{LFR} \sum_{i \in LFR} \Psi_i + \left(\frac{P_{max}^{LFR} - P_{idle}^{LFR}}{\Omega_{max}^{LFR}}\right) \left(\sum_{i \in LFR} \lambda_i\right) \right] \quad (10)$$

$$+ \left[ \tau P_{idle}^{LFS} \sum_{i \in LFS} \Psi_i + \left(\frac{P_{max}^{LFS} - P_{idle}^{LFS}}{\Omega_{max}^{LFS}}\right) \left(\sum_{i \in LFS} \lambda_i\right) \right]$$

Equation (9) shows evaluates the processing power consumption of the LF server. Equation (10) determines the networking power consumption which is composed of the power consumption of the routers (LFR) and switches (LFS) of the LF network.

**4) The total power consumption of the NF** ($\text{TPC}^{NF}$), which consists of the processing power consumption of NF processor and is given as:

$$\text{TPC}^{NF} = \left[ P_{idle}^{NF} \sum_{d \in NF} \delta_d + \left(\frac{P_{max}^{NF} - P_{idle}^{NF}}{\Omega_{max}^{NF}}\right) \left(\sum_{s \in SN} \sum_{d \in NF} X_{sd}\right) \right]. \quad (11)$$

It is worth mentioning that for NF, neither PUE nor networking power consumption is considered. This is attributed to the architecture of the NF as we assume that the processor is a Raspberry Pi board attached to an outdoor ONU. Hence, networking the incoming traffic to NF processor will be handled by the ONU and the networking power consumption of ONU will be calculated using Equation (20). Note that the ONU is assumed to be dedicated to the considered application. Therefore, the 6% fraction of ONU idle power was not included in Equation (11).

**5) The total power consumption of the VN** ($\text{TPC}^{VN}$), which is composed of the processing power consumption ($PPC^{VN}$) and the networking power consumption ($NPC^{VN}$), and is given as:

$$\text{TPC}^{VN} = (PPC^{VN} + NPC^{VN}) \quad (12)$$

where

$$PPC^{VN} = \left(\frac{P_{max}^{VN} - P_{idle}^{VN}}{\Omega_{max}^{VN}}\right) \left(\sum_{s \in SN} \sum_{d \in VN} X_{sd}\right) \quad (13)$$

$$NPC^{VN} = P_{idle}^{VW} \sum_{d \in VN} \delta_d + \left(\frac{P_{max}^{VW} - P_{idle}^{VW}}{\Omega_{max}^{VW}}\right) \left(\sum_{s \in SN} \sum_{d \in VN} \lambda_{sd}\right) \quad (14)$$

Equation (13) shows the calculation of the processing power consumption of the VN processor. Note that the idle power of the VN power profile is not represented in this equation. This is because the vehicles may spend short time periods in a car park, at a charging station or at road intersections. Therfore, processors have to be ON to utlize their processing capability immediatley as soon as a vehicle has stopped (at the intersection or in other car parks, more generally). Hence, allocating workload to the VN processor will not consume extra power as a result of activating the processor. Equation (14) describes the networking power consumption of the VN wireless adapter (VW). This adapter is assumed to be dedicated for the traffic required by the IoT source nodes. Hence, allocating traffic to this adapter will consume extra power ($P_{idle}^{VW}$) as a result of activating the adapter. Similar to the ONU, the VN WiFi adapter is assumed to be installed in the VN and dedicated for the generated task traffic. Hence, activating the wireless adapter will consume the full idle power.

**6) The total power consumption of the infrastructure network** ($TPC^{NET}$), which is composed of the power consumption of core routers ($TPC^{RR}$), metro router ($TPC^{MR}$), metro aggregation switch ($TPC^{MS}$), OLT ($TPC^O$), ONU ($TPC^U$), and AP ($TPC^A$), and given as:

$$TPC^{NET} = (TPC^{RR} + TPC^{MR} + TPC^{MS} + TPC^O + TPC^U + TPC^A) \quad (15)$$

where

$$TPC^{RR} = \rho^{NET} \left[ \tau P_{idle}^{RR} \sum_{i \in RR} \Psi_i + \left( \frac{P_{max}^{RR} - P_{idle}^{RR}}{\Omega_{max}^{RR}} \right) \left( \sum_{i \in RR} \lambda_i \right) \right] \quad (16)$$

$$TPC^{MR} = \rho^{NET} \left[ \tau P_{idle}^{MR} \sum_{i \in MR} \Psi_i + \left( \frac{P_{max}^{MR} - P_{idle}^{MR}}{\Omega_{max}^{MR}} \right) \left( \sum_{i \in MR} \lambda_i \right) \right] \quad (17)$$

$$TPC^{MS} = \rho^{NET} \left[ \tau P_{idle}^{MS} \sum_{i \in MS} \Psi_i + \left( \frac{P_{max}^{MS} - P_{idle}^{MS}}{\Omega_{max}^{MS}} \right) \left( \sum_{i \in MS} \lambda_i \right) \right] \quad (18)$$

$$TPC^O = \rho^{NET} \left[ \tau P_{idle}^O \sum_{i \in O} \Psi_i + \left( \frac{P_{max}^O - P_{idle}^O}{\Omega_{max}^O} \right) \left( \sum_{i \in O} \lambda_i \right) \right] \quad (19)$$

$$TPC^U = \left[ P_{idle}^U \sum_{i \in U} \Psi_i + \left( \frac{P_{max}^U - P_{idle}^U}{\Omega_{max}^U} \right) \left( \sum_{i \in U} \lambda_i \right) \right] \quad (20)$$

$$TPC^A = \left[ P_{idle}^A \sum_{i \in A} \Psi_i + \left( \frac{P_{max}^A - P_{idle}^A}{\Omega_{max}^A} \right) \left( \sum_{i \in A} \lambda_i \right) \right] \quad (21)$$

The PUE of the network devices, $\rho^{NET}$, in Equations (16)–(19), defines the added power consumption of network devices such as core routers, metro routers, metro switches, and OLTs attributed to cooling and lighting typically. A PUE=1 is considered for ONU and AP as both are small outdoor devices and no additional cooling installation is required for both. In addition, the 6% fraction of the idle power is not calculated for the AP as it is assumed to be dedicated for the proposed architecture and the application considered.

The objective of the model is defined as follows:

**Objective:**

Minimize the total power consumption of all processing nodes and their interconnecting networks and the infrastructure network devices, given as:

$$TPC^{CC} + TPC^{MF} + TPC^{LF} + TPC^{NF} + TPC^{VN} + TPC^{NET} \quad (22)$$

**Subject to the following constraints:**

$$\sum_{\substack{j \in NN_i \\ i \neq j}} \lambda_{ij}^{sd} - \sum_{\substack{j \in NN_i \\ i \neq j}} \lambda_{ji}^{sd} = \begin{cases} \lambda_{sd} & if\ i = s \\ -\lambda_{sd} & if\ i = d \\ 0 & otherwise \end{cases} \quad (23)$$

$$\forall\ s \in SN, d \in PN, i, j \in N.$$

Constraint (23), ensures that the total incoming traffic is equal to the total outgoing traffic for all nodes in the network, excluding the source and destination nodes.

$$\sum_{d \in PN} X_{sd} = \omega_s \quad \forall\ s \in SN \quad (24)$$

$$X_{sd} \geq \delta_{sd} \quad \forall\ s \in SN, d \in PN \quad (25)$$

$$X_{sd} \leq M1\ \delta_{sd} \quad \forall\ s \in SN, d \in PN \quad (26)$$

$$\sum_{s \in SN} \delta_{sd} \geq \delta_d \quad \forall\ d \in PN \quad (27)$$

$$\sum_{s \in SN} \delta_{sd} \leq M2\ \delta_d \quad \forall\ d \in PN. \quad (28)$$

Constraint (24) ensures that the total processing workload sent from source node $s$, allocated to a processing node $d$ is equal to the workload demand $\omega_s$ generated from source node $s$. Constraints (25) and (26) are used in the conversion of $X_{sd}$ to its equivalent binary variable. When $\delta_{sd} = 1$, the task generated from source node $s$ is allocated to processing node $d$. Constraints (27) and (28) are used to ensure that the binary variable $\delta_d = 1$ if processing node $d$ is allocated any processing workload.

$$\sum_{s \in SN} \sum_{d \in PN} \sum_{i \in NN_j} \lambda_{ij}^{sd} = \lambda_j \quad \forall\ j \in N \quad (29)$$

$$\lambda_i \geq \Psi_i \quad \forall\ i \in N \quad (30)$$

$$\lambda_i \leq M3\ \Psi_i \quad \forall\ i \in N \quad (31)$$

Constraint (29) calculates the total aggregated traffic traversing node $j \in N$. Constraints (30) and (31) are used in the conversion of $\lambda_i$ into its equivalent binary variable. When $\beta_i = 1$, the node $i$ is activated and traffic is travels through this node.

$$\lambda_{sd} = F_s\ X_{sd} \quad \forall\ s \in SN, d \in PN. \quad (32)$$

Constraint (32), ensures that the traffic from source node $s$ to processing node $d$ is equal to the data rate of the workload generated from source $s$, where $F_s$ is the ratio of the traffic to processing workload of the demand generated from source node $s$.

$$\sum_{s \in SN} X_{sd} \leq C_d \quad \forall\ d \in PN \quad (33)$$

Constraint (33) ensures that each demand generated from source node $s$ allocated to a processing node $d$ does not exceed the processing capacity of this processing node $d$.

$$\sum_{s \in SN} \sum_{d \in PN} \lambda_{ij}^{sd} \leq L_{ij} \quad \forall\ i \in N, j \in NN_i, i \neq j \quad (34)$$

Constraint (34), ensures that the traffic generated from source $s$ to processing node $d$ does not exceed the capacity of the link between any two nodes $(i, j)$.

$$\sum_{s \in SN} \sum_{d \in PN} \sum_{j \in NN_i \cap VN} \lambda_{ij}^{sd} \leq \Omega_{max}^A \quad \forall\ i \in A \quad (35)$$

Constraint (35) ensures that the total traffic traversing AP does not exceeds the capacity of the AP.

$$\sum_{d \in PN} \delta_{sd} \leq \upsilon \quad \forall\ s \in SN \quad (36)$$

Constraint (36) ensures that the processing task is not split. This may be essential in real time applications where there is no time to assemble partial results from partial processing locations. Removing this equation (instead of setting the right-hand side of the equation to $\upsilon$ splits) allows the optimization to select the best number of splits to minimize the total power consumption.

This is the other extreme compared to no splitting. Future work can consider partial splitting (different values of $v$) and hence inter processors communication where parts of the tasks are processed.

The processing allocation problem considers different evaluations with the availability of some or all processing nodes. For the VEC, multiple cases are considered (as well) to capture different vehicle densities. As mentioned previously, these VNs are clustered in a car park, at a charging station or by a road intersection within the coverage of an AP. Each vehicle is equipped with an OBU which defines the processing capability of the vehicle, and a wireless communication adapter to communicate with the AP. All VNs need to communicate with the AP, as no direct communication is allowed between VNs. All VNs are assumed to be homogeneous with the same processor capabilities. The vehicles thus work as a service provider for some of the collected data from applications related to IoT source nodes (SNs). These application tasks range from small-scale applications (with low demands), which do not require much processing capacity, to large-scale applications (high demands), which require a powerful processing node and sufficient communication link capacities to send and process the tasks. Additionally, various cases were evaluated with different ratios between processing demand and data rate demand to cover a wider range of applications. However, we assume that the tasks generated in each instance have the same processing and data rate requirements.

The allocation process of any generated task follows six phases:

**1-** A task is generated by a nearby IoT SN and is sent to an AP located in the same geographical zone. The AP has full knowledge of the available resources. This task includes the required processing and data rate per task.

**2-** The AP sends a positive acknowledgment to the task SN.

These two phases are not evaluated in the model, as they are considered control signals and, therefore, generate negligible traffic that consumes little power.

**3-** The data to be processed are sent from the SN to the AP through the wireless channel. This phase is also not considered in the model, as it is a common phase for all tasks and will not affect the task allocation decision.

**4-** The data to be processed are offloaded from the AP to one or more of the available PNs (VN, NF, LF, MF, CC). As this phase carries the main data, it will affect the power consumption and the task allocation decision. Therefore, it is treated as the main component of the model.

**5-** The extracted knowledge resulting from the processed data is sent back from the processing PNs to the AP.

**6-** The extracted knowledge is reassembled and sent from the AP to the source node that requested the service.

As we assume that the extracted knowledge has a small volume compared to the main data, the last two phases are not included in the optimization model.

## V. INPUT PARAMETERS

The proposed MILP model was evaluated using the network proposed in Figure 2. This section explains and summarizes the input data considered for the model, including the processing node capacities and efficiencies, the capacities of network devices and the power efficiencies, PUE, link capacities, and generated workload.

### A. Idle Power Consumption

Accurate values for the idle power consumption of each device are not easy to obtain in data sheets all the time. Accordingly, we based our idle power consumption values on multiple frameworks driven from the literature. First, according to [45], 90% of the maximum power consumption for network devices (routers and switches) is attributed to idle power. Therefore, this figure is used in describing the network device's power consumption in the present work. Second, based on [47], a processing node consumes 60% of the maximum power. Thirdly, in high-power-capacity network equipment, IoT applications account for a small portion of the idle power consumption. Therefore, it would be unjust to ascribe the total power consumption to a specific application. For this reason, we elected, instead, to use the 2017–2022 Cisco Visual Networking Index (VNI) [1] to express the traffic of IoT applications as a fraction of the total traffic in smart cities, such as the smart city scenario in this work. In particular, the work in [1] reported that, by 2022, IoT traffic will constitute around 6% of all global IP traffic on the Internet. Therefore, we used this number (6%) to attribute part of the idle power to these types of applications.

### B. PUE

The power usage effectiveness (PUE) is a factor used to measure the power efficiency of any network or data centre. It estimates how much power is used for the actual computing and communication, in relation to the total power resulting from computing and communication equipment plus non-IT equipment such as cooling, lighting, ventilation, etc. The PUE values of the network infrastructure and each processing node are listed in Table I.

According to Google's report in [46], PUE values have an inverse relationship to the "Space Type" of the data centre. PUE decreases with increase in the data centre size and geographical location. We assumed that PUE, at any layer, is indicative of both structure (network equipment) and function (processing). As the central cloud (CC) is a large data centre that uses sophisticated liquid and air cooling, it typically has lower PUE values compared to other processing nodes.

The telecom infrastructure is owned typically by a carrier (e.g. BT Openreach). This may give the telecom infrastructure one PUE. On the other hand, the cloud infrastructure is typically owned and operated by a cloud services provider (e.g. Amazon), which may or may not share the same building with the telecom carrier (BT Openreach), depending typically on the size of the cloud service provider and the number of racks and servers they own. This may result in a different PUE for the network devices and fog processing nodes (for example, metro switch and metro fog).

## TABLE I
### PUE VALUES FOR THE NETWORK DEVICES AND PROCESSING NODES

| Processing node | PUE value |
|---|---|
| Central cloud PUE ($\rho^{CC}$) | 1.1 [46] |
| Metro fog PUE ($\rho^{MF}$) | 1.4 [46] |
| OLT fog PUE ($\rho^{LF}$) | 1.5 [46] |
| Network devices PUE ($\rho^{NET}$) | 1.5 [18] |

### C. Capacity and power consumption of network devices

In our evaluation, we have chosen our processing devices based on the fact that the top-most layer PN (CC) has the largest processing capacity and best processing power efficiency. Conversely, the bottom-most PN (VN) has the lowest capacity and lowest power efficiency. The capacities and efficiencies of the processors of the other fog nodes vary between the CC and VN. In this model, the top-layer processor is the best in terms of capacity and efficiency.

It is worth mentioning that only one CC is considered as a local data centre. This CC is assumed to have servers that are sufficient to process all of the generated tasks. The other three fog layers (MF, LF, and NF), each has only one processor (one server in each of MF/LF, and one Raspberry Pi chip in the NF). The number of vehicles in the VEC varies based on the scenario considered. However, all VNs are assumed to be homogeneous, with the same processing capability. For security purposes, we assumed that the VN processor and its wireless communication transmitter and receiver are combined in a separate "box" which is not linked to the vehicle CAN bus. In time, the entertainment or other less critical processors in the vehicle may participate, and security must be considered if these processors or the vehicle main processors (e.g. controlling engine, windows, wipers, etc.) are used. The development of security and trust frameworks for the opportunistic vehicular clouds is outside the scope of the current work, but can build on existing cloud security and vehicle security frameworks.

The capacity of each PN is defined in terms of the Instructions per Second (IPS) it can provide. Based on device datasheets, IPS is not considered when defining CPU capability. Hence, according to [48], this value is estimated using:

$$IPS = COR \times CPS \times IPC \quad (37)$$

where $COR$ is the number of cores in a processor, $CPS$ is the processor clock rate in GHz. Both values can be extracted from the server datasheets. $IPC$ is defined as the number of instructions per cycle, and is estimated for each processing node based on the fact that a high performance processor can execute four IPCs [49].

Table II lists all values used in Equation (37) with the resultant capacity, in million instructions per second (MIPS), for each PN. In addition, the same table shows the power efficiency, in W/MIPS, calculated using the linear power profile method explained earlier in Section IV.

### TABLE II
### PROCESSING NODE POWER, CAPACITY, AND EFFICIENCY PARAMETERS.

| PN | Model | P(W) | Idle(W) | Cores | GHz | IPC | MIPS | W/MIPS |
|---|---|---|---|---|---|---|---|---|
| CC | Intel Xeon E5-2680 [50] | 115 | 69 | 10 | 3.6 | 4 | 144k | 0.00032 |
| MF | Intel Xeon E5-2630 [51] | 85 | 51 | 10 | 2.2 | 4 | 88k | 0.00039 |
| LF | Intel Xeon E5-2609 [52] | 85 | 51 | 8 | 1.7 | 4 | 54.4k | 0.00063 |
| NF | RPI 4 Model B [53] | 15 | 9 | 4 | 1.5 | 1 | 6k | 0.001 |
| VN | MobiWAVE iMX6 [54] | 10 | 6 | 2 | 0.8 | 2 | 3.2k | 0.00125 |

For the other network devices, Table III shows the power consumption and capacity values in the different layers, taking into account the percentage of the idle power attributable to cloud and fog traffic. In addition, Table III shows energy efficiency values for network devices.

### TABLE III
### NETWORK DEVICES POWER, CAPACITY AND EFFICIENCY PARAMETERS.

| Network Layer | Device | P(W) | Idle (W) (90%) | Idle (W) (6%) | Gb/s | W/Gb/s |
|---|---|---|---|---|---|---|
| Core layer | Core router port [55] | 638 | 574.2 | 34.452 | 40 | 1.595 |
| Metro layer | Metro router [56] | 25 | 22.5 | 1.35 | 40 | 0.063 |
| | Metro switch [57] | 500 | 450 | 27 | 1800 | 0.028 |
| Access layer | OLT [58] | 50 | 45 | 2.7 | 1920 | 0.003 |
| | ONU [59] | 15 | 13.5 | - * | 10 | 0.150 |
| Edge layer | AP [60] | 11 | 4.8 | - * | 1.167 | 5.313 |
| CC network | CC Router port [56] | 25 | 22.5 | 1.35 | 40 | 0.063 |
| | CC Switch [61] | 460 | 414 | 24.84 | 600 | 0.077 |
| MF network | MF Router port [62] | 13 | 11.7 | 0.702 | 40 | 0.033 |
| | MF Switch [63] | 245 | 220.5 | 13.23 | 200 | 0.123 |
| LF network | LF Router port [62] | 13 | 11.7 | 0.702 | 40 | 0.033 |
| | LF Switch [63] | 245 | 220.5 | 13.23 | 200 | 0.123 |
| VN network | VN Wi-Fi adapter [64] | 2.5 | 1.5 | - * | 0.0722 | 13.850 |

* the device is assumed to be fully dedicated to the application considered

### D. Processing and data rate requirements

In our evaluation, as explained earlier, we define the CPU capacity in MIPS. Taking inspiration from [65], one of the analyzed IoT sensors was a smart city-based sensor designed to detect environmental events using an earthquake prediction algorithm. It is reported that a job with a size of 11.72 kB (0.09 Mb/s data rate) needs 78 MIPS to be processed. Through simple calculations, we derived the processing requirement. We assumed a minimum required data rate equal to 1 Mb/s, and calculated the required MIPS to execute 1 Mb of data as follows:

$$\Phi = \frac{78 MIPS}{\frac{0.09 Mb}{s}} = 866 \cong \frac{1000\ MIPS}{\frac{Mb}{s}} \quad (38)$$

Based on the above calculation, and as we assumed a futuristic increase of the traffic demands, we selected the 1,000 MIPS as the minimum required processing demand per task and examined a range of settings where we increased this demand up to 10,000 MIPS per task.

We assumed that the data rate of any task rises with increase in the processing workload. Therefore, in Equation (39), we introduce a relationship between the processing workload demand and the data rate demand for each requested task as a ratio, termed as 'data rate ratio' (DRR). A fixed DRR, equal to 0.001, is considered for the first two evaluations in Section VI (B). Therefore, the required data rate ranged from 1 Mb/s to 10 Mb/s, and thus

$$DRR = \frac{traffic\left(\frac{Mb}{s}\right)}{processing\ (MIPS)}. \qquad (39)$$

On the other hand, different DRR values were assessed in Section VI. It should be noted that small DRR value represents an (IoT) application where small data volumes are generated, for example by measuring a physical quantity, followed by extensive processing of the data. Large DRR values may represent situations where large data volumes are generated followed by limited processing, for example as in video streaming or some forms of gaming.

## VI. EVALUATION SCENARIOS AND RESULTS

### A. Scenarios considered

This section describes the scenarios and the architecture considered in the proposed processing allocation MILP model. Three main dimensions were studied, summarized as follows:
- Centralized versus distributed processing (in CC, MF, LF, and NF)
- Opportunistic processing (in VEC)
- Single versus distributed task allocation (non-splittable versus splittable demands)

Two different architectural designs, inherited from Figure 2, were assessed based on the design of the edge network. The first architecture considers one zone, with multiple VEC clusters. This design captures an urban area where many car parks, charging stations or road intersections are available. It also depicts VEC clusters, each consisting of VNs parked in a car park, at a charging station or stopping at a road intersection and connected to an AP located in the same cluster, as illustrated in Figure 1. With different potential locations for the source generation and with different APs in each cluster, we can assess the effect of different realistic scenarios. In the second architecture, we expanded the PON network to include multiple zones, with numerous ONUs, each with one or more AP (VEC cluster). Considering different zones in this architecture allowed us to mimic an expanded urban area (a city, for example), where the infrastructure can connect multiple VECs located in different zones within the city.

In each architecture, four cases were evaluated to capture the main task allocation dimensions, as follows:
- Central Cloud Allocation (CCA). In this case, centralized processing was evaluated by allocating all tasks to the central cloud servers. This represents a baseline approach to which other cases (considering the VEC) are compared.
- Cloud-Fog Allocation (CFA): In this case, we considered a cloud-fog architecture, with available cloud and fixed fog PNs, but with no available VNs. This case introduces processing over distributed locations. It also represents off-peak periods over the day where no vehicles are in the city car parks and charging stations (and at intersections) or situations where vehicles are not participating in the resource provisioning service.
- Cloud-Fog-VEC Allocation with low vehicular nodes density (CFVA-L): In this case, VNs are introduced at a low density. Each VEC within a cluster includes two VNs, with the total number of available VNs equal to eight. This choice limits the size of the problem to a size that can be handled efficiently by the MILP given that the allocation problem is known to be NP hard. This case represents low peak periods with a limited number of vehicles.
- Cloud-Fog-VEC Allocation with high vehicular nodes density (CFVA-H): In this case, the number of VNs is increased to 15 VNs per VEC, with a total of 60 available VNs over four VEC clusters/zones. This case represents a high density of VNs, as experienced in the peak periods during the day.

The last two cases, CFVA-L and CFVA-H are assessed with single and distributed (non-splittable and splittable) allocation to capture different real-time demands. The two architectures and their results are explained next in Section B. As the evaluations in these two sections considered a fixed DRR value, with DRR=0.001, Section C presents a comparison study at different DRR values, to evaluate the impact of the different processing and traffic volume on the processing allocation and the total power consumption

### B. Power Consumption Evaluation

#### 1) Processing Allocation in Cloud-Fog-VEC Architecture with One Zone and Multiple VEC Clusters

The end-to-end architecture considered in this evaluation is similar to the architecture presented in Figure 2. However, the edge network evaluated in this section includes only one zone (with one ONU). This zone consists of one ONU (as mentioned above) and four clusters. Each cluster represents a car park, a charging station or road intersection with an AP and connected VNs. Tasks are generated from nearby IoT SNs located in the same cluster. These SNs are assumed to have connection only with their local AP and cannot communicate with any AP located in a different cluster. Therefore, different tasks can be allocated by each AP. In this section, four scenarios are defined to capture different situations based on the number of generated tasks and the location of the source node. In all evaluated scenarios, the tasks processing requirements ranged from 1,000 MIPS to 10,000 MIPS. The required data rate for the tasks increased, based on DRR=0.001, from 1 Mb/s to 10 Mb/s. Furthermore, four cases were assessed, as explained in earlier, referred to as Central Cloud Allocation (CCA), Cloud-Fog Allocation (CFA), Cloud-Fog-VEC Allocation with Low VN (CFVA-L) and Cloud-Fog-VEC Allocation with high VN (CFVA-H), both with single (SA) and distributed (DA) allocation strategies.

**SCENARIO#1. One task generated from one cluster**
In this scenario, we assumed that, in any time instance, one task is generated from a source node located in one of the four clusters. Figure 4 shows the total power consumption of the different cases considered. As the power consumed is a function of the networking and processing power consumption, the processing placement becomes a result of a trade-off between networking and processing power values to achieve the optimum processing allocation. This optimized allocation at each processing node (PN) of the CC, Metro Fog (MF), OLT Fog (LF), ONU Fog (NF), and Vehicular Edge Cloud (VEC) is summarized in Figure 5.

Figure 4 shows that allocating all tasks to CC consumes the highest power with a linear increase in relation to the size of the generated tasks. Given the absence of VNs in CFA, tasks are allocated to the most efficient fixed PN, until it is fully exhausted or becomes too thin to accommodate the task. For example, with low workload demands, NF is the most efficient PN, as seen in Figure 5. This is because it is closer to the edge and therefore consumes less networking power, achieving up to 87% power saving, compared to CAA. However, at 7000 MIPS, NF becomes too thin to allocate tasks with high demands. Hence, tasks are allocated to the next most efficient fixed fog node (LF). This explains the jump in the power consumption where the power saving is reduced to 44%, (also compared to CCA).

When vehicular nodes become available, both single and distributed allocation strategies were assessed with low and high VN density. In the case of single allocation (CFVA-L and CFVA-H), and as shown in Figure 4, the model achieves an early power saving of, on average, 70%, compared to CFA. This saving is caused by allocating the tasks to the available VN (as in Figure 5) thus, avoiding the activation of the ONU and its fog server. However, this saving is limited to cases where demands are within the VN processor's capacity. This explains the increase in power consumption, at 4000 MIPS, where all tasks are allocated to NF, and therefore in this case the power consumption is the same as that of the CFA. It is also observed that, in single allocation, the VN density has no effect on the allocation decision, and therefore, has no influence on the power consumption. This is because, with the no-splitting constraint, only one VN is needed to serve the generated task.

In the distributed allocation case, Figure 4 shows that, with low demands, the same power saving as that seen in the single allocation is achieved. This is because one VN was enough to serve the generated task, and therefore the splitting flexibility was not needed. However, a continuous power saving is observed regardless of the VN capacity limitation (i.e. beyond the 3000 MIPS), achieving up to 71% power saving, in comparison to single allocation. This is attributed to the splitting ability of the model and bin-backing the split task into the available VNs, therefore achieving better power saving. Moreover, it was observed that when the available VNs are not sufficient to serve the full task, the model initially allocates a major part of the task to the most efficient fixed PN until it is fully utilized, and then allocates the remaining part of the task to the available VN. For instance, in Figure 5 (in CFVA-L and at 7000 MIPS), it can be seen that the available VNs (with a total of 6400 MIPS capacity) are not able to serve the whole task. Hence, the optimization outcome activates NF with 6000 MIPS allocation, and then allocates the remaining sub-task to the local VEC. This causes the majority of the task to be allocated to NF, as it has a more efficient processor than the VN. In contrast with single allocation, VN density has an impact on the task allocation and the power consumption. The high density VN increased the capacity of the VEC and, therefore, more tasks were allocated to the available VNs and ONU activation was avoided. Thus, further power saving of up to 50% were achieved, compared to the low VN density (CFVA-L (DA)).

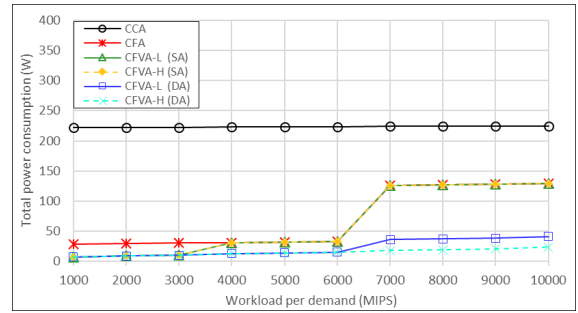

Fig. 4. Total power consumption, in Scenario 1 with one zone.

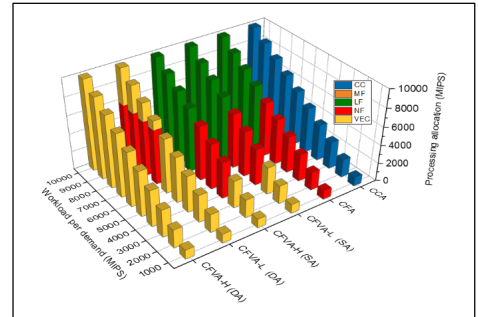

Fig. 5. Processing allocation in each PN, in Scenario 1 with one zone.

Part of the aim of this work is to design a VEC architecture that is able to expand and connect many VEC clusters together in a cloud-supported architecture. Accordingly, it is very important to study the allocation behaviour among the VEC clusters considered. Figure 6 summarizes the processing allocation in each individual VEC, taking into account the fact that the task was generated from VEC1 (this will be referred to as local VEC).

The single allocation results show that, regardless of the VNs' density, the tasks allocated to VECs were limited to the 1000–3000 MIPS range. This is due to the VN limited capacity. Hence, all tasks were allocated to one VN in the local VEC, as the model was constrained by the "no splitting" condition. Similarly, in distributed allocation, a single task with high demand was split and allocated to VNs located in the local VEC. No splittable tasks were allocated to the other VEC clusters, even when the VNs in the local VEC were exhausted. This is because allocating sub-tasks to a non-local VEC will activate another ONU. Hence, it is more efficient to utilize the NF, with its efficient processor, rather than allocating the sub-task to a non-local VEC. On the other hand, the same figure shows that, as the VNs' density increased to 15 VNs per VEC, the local VEC became enough to allocate the whole task, even with the increasing processing demand.

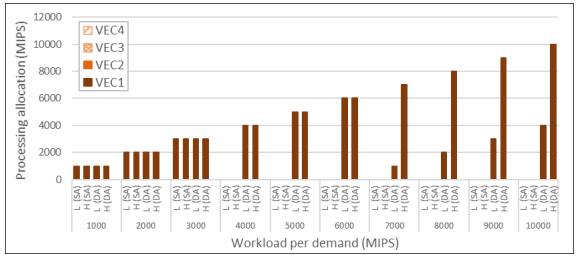

Fig. 6. Processing allocation in each VEC in CFVA (SA) and CFVA (DA), in Scenario 1 with one zone.

**SCENARIO#2. One task generated from each cluster.**
In this scenario, we investigate the impact of increase in the number of the tasks and the activation of multiple APs, by generating one task from each cluster with a total of four tasks. Figures 7 and 8 show the total power consumption and the processing allocation in each PN, respectively.

In this scenario, we observed less power saving, as shown in Figure 7, compared to Scenario 1. This is attributed to the increase in the generated tasks. In addition, Figure 8 shows an allocation behavior which is relatively comparable to scenario 1. The bottom-most processing nodes have the most efficient total power consumption due to their associated low networking power consumption. Hence, tasks are allocated first to these nodes if they can satisfy the processing workload. Moreover, when the most efficient PN cannot accommodate all of the generated tasks, an upper PN becomes the most efficient location, and is fully utilized first. For example, in CFA, with 2000 MIPS or more, NF cannot allocate all of the four tasks (with a total of 8000 MIPS), as this exceeds the NF capacity. Although NF can accommodate three out of the four tasks, all tasks were allocated to the LF, as seen in Figure 8. This behavior is attributed to two factors. First, activating one PN is more efficient than activating both NF and LF. This is due to the power overhead and idle power resulting from activating two PNs. Second, once the OLT and its fog server are activated, it is more efficient to allocate all tasks to this PN, as its processor has better efficiency compared to the NF processor. Despite the NF limitation, allocating tasks to LF yields up to 41% power saving, compared to CCA.

Similar to Scenario 1, VN density has no effect on the single allocation cases (CFVA-L and CFVA-H), as the number of generated tasks is small enough to be satisfied by the available VNs when the demands of the (four tasks) are within the VN processor's capacity. However, in CFVA-L(SA), we observed an allocation in LF; specifically, when the VNs were exhausted. The tasks were allocated to LF rather than NF. This is, again, due to the same reason of avoiding the overhead resulting from activating two PNs. This explains the jump in the power resulting at 4000 MIPS, as seen in Figure 7. Moreover, CFVA-L and CFVA-H, with distributed allocation, have comparable allocation results to Scenario 1, except that with increasing demands beyond 7000 MIPS. Here, all tasks were allocated to the LF, as both NF and VEC combined cannot satisfy the required demands. Despite the small power saving achieved in this scenario, CFVA (in both allocation strategies and with both VN densities) can still save power by up to 87% and 61%, compared to CCA and CFA respectively. Moreover, splitting tasks in distributed allocation guarantees a continuous power saving, with high demands, of, on average, 59%, compared to single allocation.

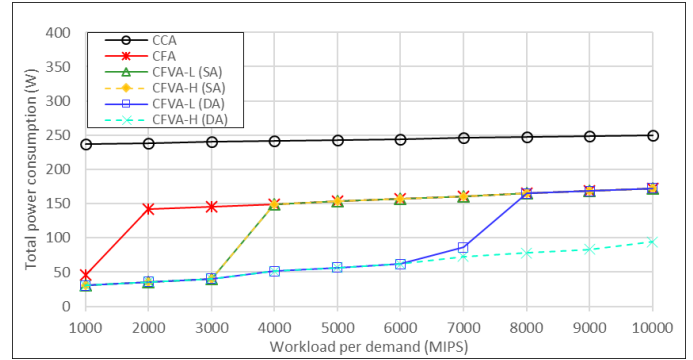

Fig. 7. Total power consumption, in Scenario 2 with one zone.

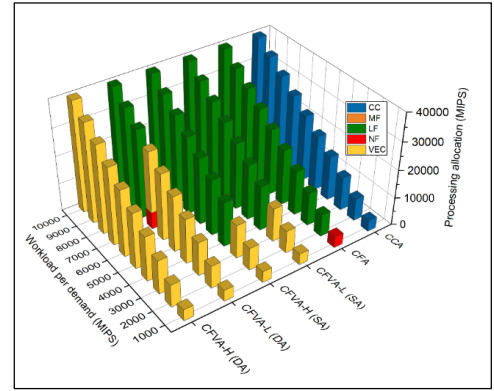

Fig. 8. Processing allocation in each PN, in Scenario 2 with one zone.

Figure 9 summarizes the processing allocated in each VEC in this scenario. Recalling that each task was generated from a different cluster, an equal utilization is observed, for all VEC clusters, as each VEC is considered a local processing node for one task. Despite this, in DA, and by 7000 MIPS, the VEC was unequally utilized, as the insufficient capacity of the available VNs caused the activation of the ONU and NF processor. Thus, it was more efficient to fully utilize the NF and then allocate the remaining processing from each task to its local VEC.

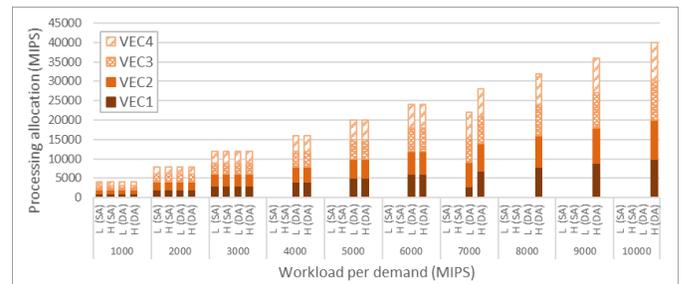

Fig. 9. Processing allocation in each VEC in CFVA (SA) and CFVA (DA), in Scenario 2 with one zone.

**SCENARIO 3. Five tasks generated from one cluster**
In this scenario, the number of generated tasks has increased to five tasks originating from one cluster. The allocation trend in this scenario follows the same trend observed in Scenarios 1 and 2, highlighted through the following two points. First, the PN nearest to the SN is the most efficient in total power consumption. Second, when the most efficient PN cannot satisfy all the required demands, an upper layer PN is activated and is fully utilized before utilizing other PNs. Following this, the optimization begins to allocate the remaining tasks (or splittable tasks), if there are any, to the nearest PN, which has enough capacity to accept the allocation. The results supported by these trends are shown in Figures 10 and 11.

The results illustrate that, despite the increase in the number of generated tasks, the VEC is still attractive, and efficient, when it comes to serving the generated tasks. Moreover, unlike the previous two scenarios, increasing the VNs' density affected the single allocation (SA) results in CFVA-H, with up to 42% power saving compared to CFVA-L. This is because, with low density, the local VEC was not enough to serve the five tasks generated (with 2000 and 3000 MIPS). Thus, allocating tasks to NF was more efficient than sending these tasks to a non-local VEC, as observed in Scenario 1. Moreover, increasing the number of VNs in each cluster, in CFVA-H (SA), enables the optimization to allocate all generated tasks to the local VEC, whenever this is sufficient.

In distributed allocation (DA), CFVA-L shows an early increase in the power consumption, as seen in Figure 10. This is attributed to a new allocation behavior where the optimization allocates tasks to a non-local VEC, as seen in Figure 12 (at 3000–6000 MIPS). The non-local VEC in this case became an efficient location combined with the NF and local VEC. Moreover, utilizing these two locations (NF and VEC) is more efficient than activating the OLT and its fog server (LF). However, this allocation causes an increase in the power consumption, which explains the early and continuous power saving in CFVA-H(DA) with 34%–48% power saving compared CFVA-L (DA).

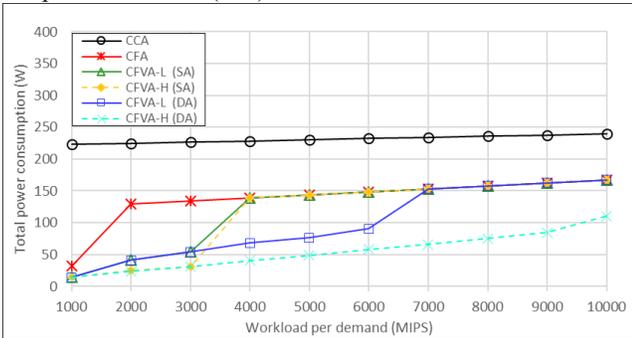

Fig. 10. Total power consumption, in Scenario 3 with one zone.

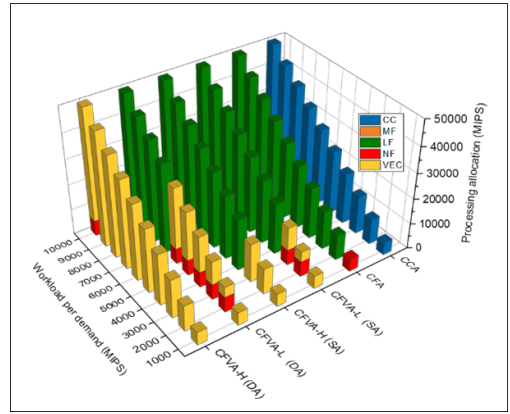

Fig. 11. Processing allocation in each PN, in Scenario 3 with one zone.

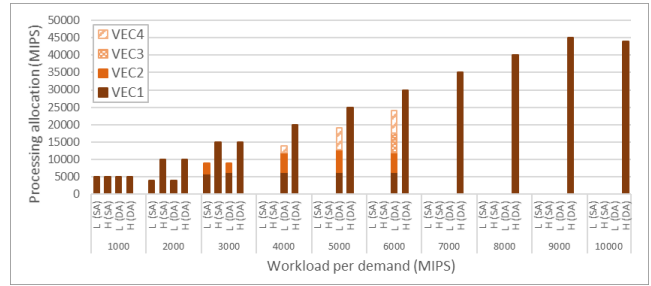

Fig. 12. Processing allocation in each VEC in CFVA (SA) and CFVA (DA), in Scenario 3 with one zone.

## SCENARIO 4. Five tasks generated from each cluster

In this scenario, we further increased the number of generated tasks and assumed that each cluster received five generated tasks from a local source node.

As seen in Figure 13, CCA experiences a nonlinear increase in the power consumption. This is due to the extra server(s) activated to accommodate the increased demands. It is also observed that the power savings in CFA and CFVA experience a dramatic reduction, compared to the previous scenarios. This is also attributed to the increased number of generated tasks and the elevated number of activated PNs, as shown in Figure 14.

In CFA, the model still yields a power saving of 20%–37% with the low demands, compared to CCA. This occurs when tasks are allocated to NF, LF, and MF (Figure 14). However, this saving stopped by 5000 MIPS when all tasks were allocated to the highest processing location, i.e. CC. The reason for this allocation is attributed to two factors. First, the high-demand (in terms of number of MIPS needed) tasks cannot be satisfied by the access layer PNs, which are the most efficient location (i.e. LF and NF). Second, although MF and LF servers, combined, can serve the required generated workload, it is more efficient to activate one server in the CC instead of activating two MF and LF servers. For example, activating MF and LF consumes a combined idle power of 102 W. In contrast, activating one server in the CC consumes only 69 W. Moreover, the CC has a very low PUE of 1.1, compared to 1.4 and 1.5 PUE in MF and LF, respectively. As a result, CC becomes the optimum placement to process the generated tasks.

In CFVA-L (SA), similar behavior is observed with less VEC utilization and early activation for the CC server. In CFVA-H (SA), the increase of VNs density (with low demands) has a clearer effect on the power consumption, compared to the low density case. As seen in Figure 15, a 36%–42% power saving is achieved by CFVA-H (SA) compared to CFVA-L (SA). Although this power saving is limited to two cases with low demands, this can show that the total capacity of the available VNs processor is crucial in the allocation decision. The impact is confirmed in the distributed allocation (CFVA-L and CFVA-D), where more tasks are allocated to the VEC, as seen in Figure 16. Moreover, the splitting and bin-packing flexibility of the model increase the VEC utilization, and show the impact on the power saving.

Despite the common behavior of allocating the high-demand tasks to CC, one feature was noticed, namely that in some cases the optimization allocated a portion of processing to VEC. For instance, in CFVA-L (DA) at 7000 MIPS, the MILP solution allocated all tasks to CC, but at 8000 MIPS, a portion was allocated to the VEC. This can be justified through two observations. The first is the common behavior of selecting the PN with the smallest spare capacity that is enough to allocate the tasks, and then allocating the remaining portion to the bottom most sufficient PN (VEC in this case). The second reason is that activating each server in CC consumes a power overhead resulting from the idle and PUE values. Thus, it is more efficient to allocate the remaining processing (16000 MIPS) to VEC rather than activating a new CC server.

Figure 15, shows that both VNs' density and the distributed allocation strategy have a substantial impact on the processing allocation in VEC. Moreover, the source of the generated demands is another factor which helps to maintain an equal utilization for all VEC clusters, despite the increase in the processing demands.

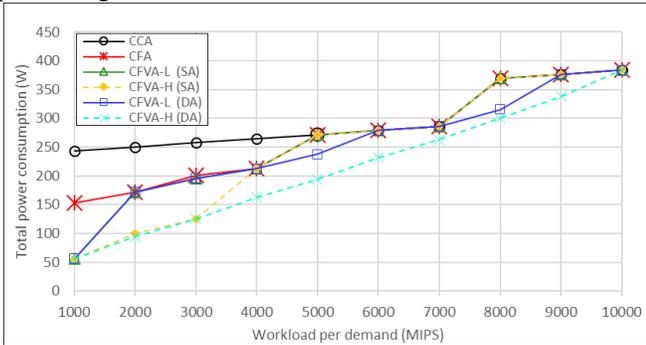

Fig. 13. Total power consumption in Scenario 4, with one zone.

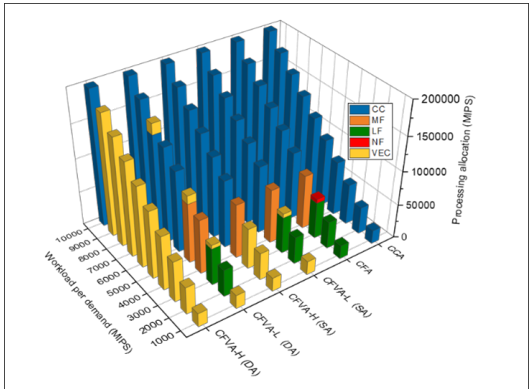

Fig. 14. Processing allocation in each PN, in Scenario 4, with one zone.

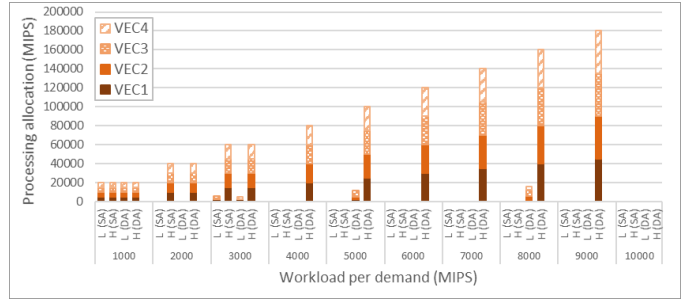

Fig. 15. Processing allocation in each VEC in CFVA (SA) and CFVA (DA), in Scenario 4 with one zone.

### 2) Processing Allocation in Cloud-Fog-VEC Architecture with Multiple Zone

In this section, similar to the previous section, we based our evaluation on the architecture shown in Figure 2. However, instead of having one zone in the edge network, we considered four zones representing different geographical areas, illustrated in Figure 16. The access network has also expanded, with four ONUs, each allocated a fog unit (NF) with a total of four NF nodes (instead of one NF node in the previous Section (1)). To reduce the model's complexity, i.e. its runtime, and to be able to examine scenarios with high VNs density, we considered a setting where each zone includes only one cluster (with one VEC and one AP). The aim of this expanded design is to evaluate the processing allocation with multiple VEC clusters located in different zones. Similar to the evaluation scenarios described in Section (1), four scenarios were considered, with the same highlighted cases and tasks requirements.

**SCENARIO 1. One task generated from one zone**

In this scenario, it is assumed that one task is generated from SNs located in one of the connected zones. Figures 16 – 18 show identical power consumption and allocation results compared to the same scenario with one zone. The optimum solution allocates one generated task to the same optimum locations observed in Scenario 1 with one zone, for all evaluated cases. This is because the task will be processed locally in its NF or VEC as the most efficient location, and will never be allocated to the non-local NF or VEC.

Moreover, in single allocation, tasks can be accommodated by one VN in a local VEC, as long as the VN capacity satisfies the task processing requirements. On the other hand, in distributed allocation, one task can be split between VNs located in the same VEC, but not between VECs in different clusters/zones (or different fixed PNs). This confirms that the "one task allocation" is not be affected by the edge network design.

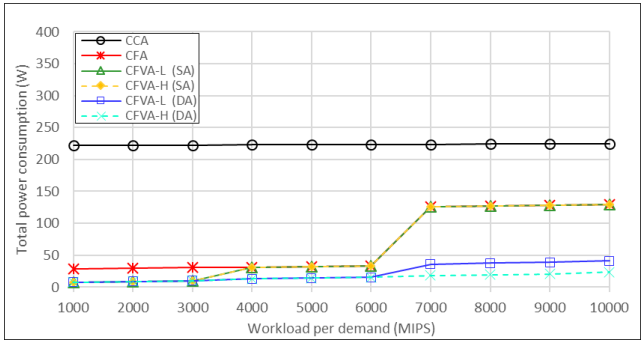

Fig. 16. Total power consumption, in Scenario 1 with multiple zones.

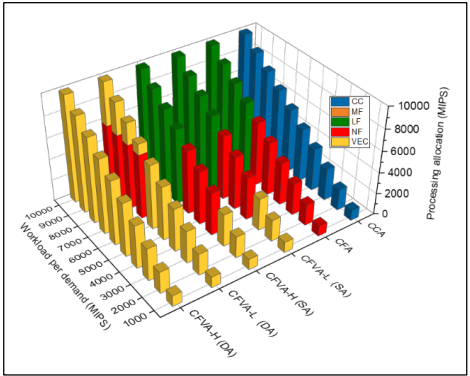

Fig. 17. Processing allocation in each PN, in Scenario1 with multiple zones.

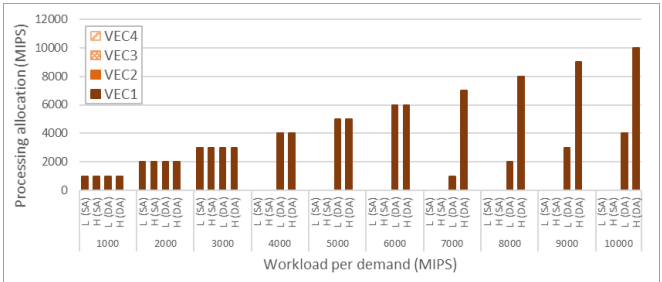

Fig. 18. Processing allocation in each VEC in CFVA (SA) and CFVA (DA), in Scenario 1 with multiple zones.

**SCENARIO 2. One task generated from each zone**

The results in Figures 19 and 20 illustrate different power trends and processing allocations compared to Scenario 2 with one zone (Figures 7 and 8).

Increasing the number of ONUs (and NF nodes) increases the power consumption in some cases and saves more power in other cases. For instance, in CCA (baseline case), the power consumption increased compared to Scenario 2 with one zone. This is because the four ONUs were activated to forward the tasks to the CC from each zone. Moreover, in CFA (in this scenario), the power consumed with 1000 MIPS (90W), in Figure 19, is more than that consumed with the same demand size in Scenario 2 with one zone (50 W), as seen in Figure 7. This is attributed, again, to activating the four ONUs; conversely, with one zone architecture, the only available ONU was activated. On the other hand, with 2000 MIPS demand, Figure 19 shows a decrease in the power consumption (103 W) compared to the power consumed with the same demand in Figure 7 (141 W). This is because, in the "one zone architecture", one NF was not enough, and so the LF was activated. In this scenario, the availability of four NF nodes avoids activating the LF and results in a lower power consumption.

It was also noticed that the local NF allocation was not always the optimal decision. For instance, with 1000 MIPS, each task generated from a zone was allocated to the local NF of zone 3. This is because activating each NF will cost an extra NF idle power of 9 W, due to the activation of a new processor (the four ONUs were activated just to forward the task to one zone). Thus, the optimization tends, if possible, to reduce the number of activated PNs by allocating all generated tasks to one PN. By 4000 MIPS, each task was allocated to the local NF, as the limited capacity of the NF processor prevented allocating more than one task to the same NF. This resulted in a small increase in the power consumption, as seen in Figure 19, since the four NF nodes were activated and allocated a task.

In single allocation (in CFVA-L and CFVA-H), Figure 19 shows identical power consumption, with low demands, as the power consumed in the same scenario with one zone. The power afterwards followed the same CFA result as NF, and then LF, were activated when the available VNs became thin. It is also confirmed in Figure 20 that the VNs' density has no effect on the allocation decision (as is the case for one zone architecture).

In distributed allocation, Figure 19 shows better power consumption in CFVA-L, compared to Figure 7 with the one zone architecture. Moreover, the overall VEC utilization was improved, as seen in Figure 20. The reason for this is that, with the increasing number of the NF nodes, the MILP was able to split the local task between the local VEC and NF, instead of activating the OLT and LF node. Accordingly, the CFVA-L (DA) with multiple zones achieves better VEC utilization compared to the one zone architecture. It was also observed in Figure 21 that all VEC clusters were equally utilized, again, a behaviour comparable to Scenario 2 with one zone (Figure 9), as each task was processed locally.

In distributed allocation with low VN density (CFVA-L), the increased number of NF nodes helps in allocating more tasks to the available VNs, as this allocation is more efficient than activating the LF in the one zone architecture (Figure 8). However, the four VEC clusters in CVFA-L(DA) (Figure 21) were not equally utilized, as this depends on the amount of processing allocated to NF, which leads the remaining workload to be allocated to the local VEC.

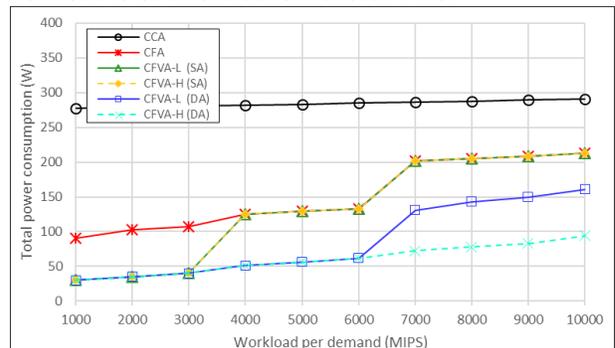

Fig. 19. Total power consumption, in Scenario 2 with multiple zones.

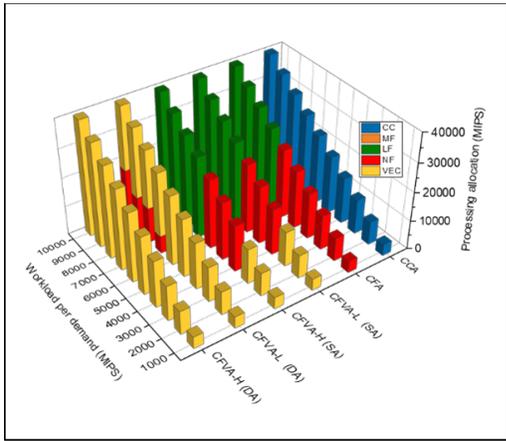

Fig. 20. Processing allocation in each PN, in Scenario 2, with multiple zones.

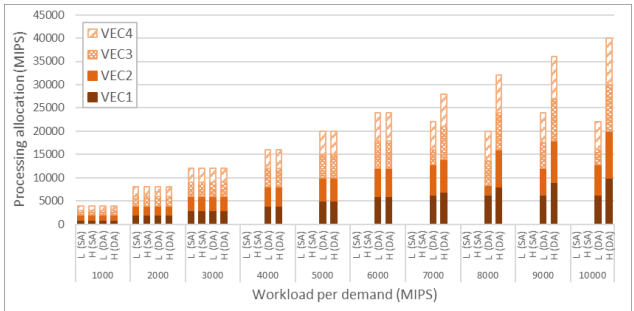

Fig. 21. Processing allocation in each VEC in CFVA (SA) and CFVA (DA), in Scenario 2 with multiple zones.

**SCENARIO 3. Five tasks generated from one zone.**

The allocation results in Figures 22 and 23 follow the same trend observed in Scenario 3 with one zone (Figures 10 and 11). The increase in the number of NF nodes reduces the power consumption whenever utilizing NF nodes leads to avoiding OLT and LF activation. For example, in CFA with 1000 MIPS, one NF was enough to serve the generated tasks in both the one and four zone architectures. Afterwards, the one available NF in the "one zone architecture" became exhausted and the tasks were allocated to LF, which consumes 129 W (Figure 10). On the other hand, in this scenario the tasks were accommodated in two NF nodes, thus consuming only 64 W (Figure 22). The power increased afterwards based on the number of activated NFs, until reaching a linear increase when LF became the most efficient PN.

The allocation behavior of the remaining cases of CFVA for single and distributed strategies followed a comparable result to that seen with the one zone architecture, summarized as follows: the local VEC is always the most efficient location. If local VEC is thin, local NF is fully utilized first, then the local VEC is utilized to accommodate the extra processing demand. If both were fully utilized, then a non-local NF is utilized before the VEC located in that same zone. However, one case was observed in CFVA-L(DA) where the optimization allocated the tasks to non-local VEC without utilizing the NF node. For example, at 5000 MIPS, the optimization fully utilized the local NF and local VEC with 6000 MIPS and 6400 MIPS allocation, respectively. The optimization allocated the remaining 12600 MIPS, i.e. the remaining tasks to two non-local VECs without activating or utilizing their local NF. This is shown in Figure 23 (at 5000 MIPS) where a small portion was allocated to the NF despite the other cases (i.e. with 4000 and 6000 MIPS). This is because the remaining workload (12600 MIPS) could be satisfied either by two NF processors and two VNs, or four VNs. As a result, activating two access points and four VNs was more efficient than activating two NFs. This is attributed to the idle power consumed by activating the NF processor (18 W for two NFs). Moreover, and based on the status of the VN in an active mode, allocating tasks to four VNs consumes only a total of 15.6 W (6 W for activating the four VN Wi-Fi adapters, and 9.6 W for activating two APs).

Regarding the VEC utilization, we noticed a decrease in the utilization (specially for the non-local VECs) with low VN density (for both single and distributed allocation), as seen in Figure 24. This change is attributed to the increase in the number of NFs, which accommodates more processing instead of the available VEC, except for some cases with high-demand split tasks (as described above). This is because allocating a task to a non-local VEC will activate the zone ONU and Cluster AP. Therefore, allocating this task to the NF saves the power consumption resulting from activating the AP (taking into account that only the local AP is assumed to be activated with the generated tasks).

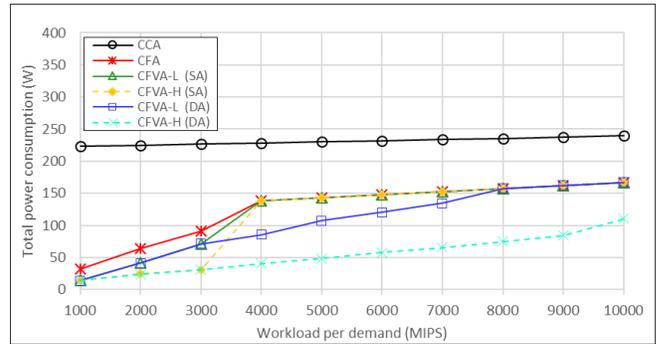

Fig. 22. Total power consumption, in Scenario 3 with multiple zones.

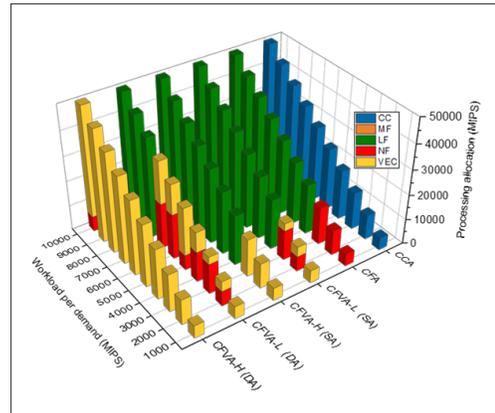

Fig. 23. Processing allocation in each PN, in Scenario 3 with multiple zones.

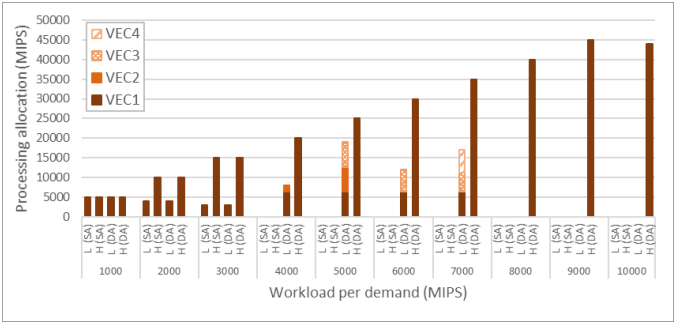

Fig. 24. Processing allocation in each VEC in CFVA (SA) and CFVA (DA), in Scenario 3 with multiple zones.

**SCENARIO#4. Five tasks generated from each zone**
This scenario assumed a total of 20 generated tasks (5 tasks from each zone). Compared to the same scenario with one zone, Figure 25 shows relatively comparable power trends for all cases, except for a slight improvement in the power saving in the CFVA-L(DA), with up to 80% and 56%, compared to CCA and CFA, respectively. This is due to the ability of the optimization, with the support of NF availability, to bin-pack small split tasks in the VEC, as observed in Figure 26. These small split allocations cause unequal utilization for the available VECs, as seen in Figure 27. This is because the model utilizes the optimum number of large PNs to be activated. For instance, in the case of CFVA-L(DA) at 9000 MIPS, the model fully utilizes one server in CC and two NF processors, following which it utilizes the VEC with remaining splits. With high density VNs, the model returns to equally utilize the VECs as the local cluster becomes efficient and sufficient to allocate all generated tasks, up to 9000 MIPS.

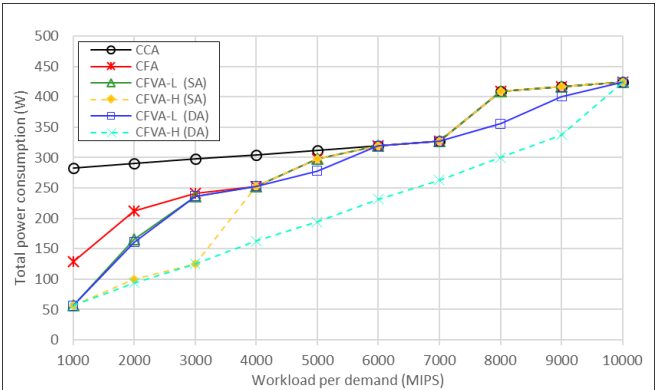

Fig. 25. Total power consumption, in Scenario 4 with multiple zones.

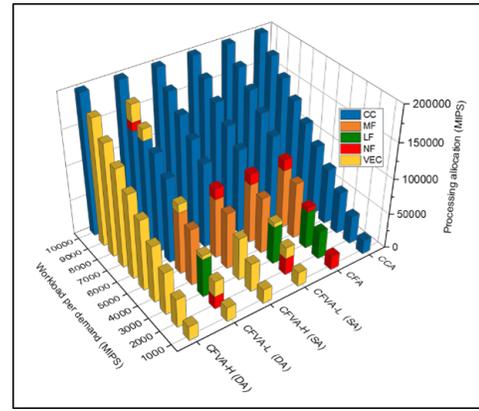

Fig. 26. Processing allocation in each PN, in Scenario 4 with multiple zones.

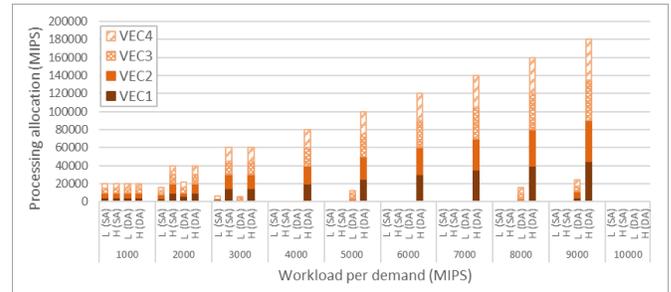

Fig. 27. Processing allocation in each VEC in CFVA (SA) and CFVA (DA), in Scenario 4 with multiple zones.

*3) The effect of demands variety on the processing allocation in Cloud-Fog-VEC Architecture with One Zone*

The evaluations in the previous two sections (1 and 2) were conducted with fixed demands requirements for the processing (MIPS) and data rate (Mb/s). As explained previously, we considered generated demands with a minimum of 1000 MIPS requirement per task. Moreover, the relation between the required processing and data rate is based on a defined value referred to as Data Rate Ratio (DRR). Thus, the increased processing demands in the previous evaluations were accompanied by a fixed increase in the required data rate based on a DRR value equal to 0.001. In this section, we aim to explore different ranges of DRR values that create varied ranges of generated demands which can represent multiple applications. The goal of exploring these ranges is to examine a vast array of input sets to assess the proposed architecture and the developed model with different types of potential applications.

We considered two sets of processing demands (high and low) to capture the needs of tasks in terms of processing workload. The high demands input set varied from 1000 MIPS to 10000 MIPS requirement per task. On the other hand, the low demand set varied between 100 and 1000 MIPS per task. Moreover, we considered increasing the required data rate per task, for both processing demand sets. Different DRR values were defined, varying between 0.001 and 0.8 to consider different scenarios with low and high data rates. These DRR values were chosen from a wide range of possible values where different behaviours of the processing allocation were observed. It is worth mentioning that results with a DRR value below 0.001 were not presented, as they proved to have the same model behaviour and allocation decisions as the results of 0.001. The same was true for values higher than 0.8, where comparable results were observed to the DRR=0.8 results. The former (DRR<0.001) indicates that traffic is very low and processing power consumption dominates the allocation decisions leading to the use of the most energy efficient processor. The converse is true at DRR>0.8, where traffic power consumption dominates, leading to the use of the nearest processor.

The following are the values considered for DRR: 0.001, 0.04, 0.08, 0.1, 0.2, 0.4, and 0.8. As examples of the types of applications represented by these DRR values, a DRR value of 0.001 represents a task that is intensive in processing and light in communication, for example sensing simple data and then processing it intensively. At the other extreme, the DRR of 0.8 may represent processing video streams, which is intensive in communication and processing. Other applications could include video streaming, images or large sensor files.

In this evaluation, we considered end-to-end architecture, where the edge network consists of one zone and four VEC clusters. We evaluated the scenario where one task is generated from one cluster. It is worth mentioning that, as we challenge the network with high DRR values, and therefore high traffic, the source node generating the task is assumed to have a wired connection to the AP to accommodate the large amount of generated traffic. For example, with a processing demand of 1000 MIPS at DRR of 0.8, the generated traffic is equal to 800 Mb/s, which can be sent through the wireless 1 Gb/s connection of the AP. However, with a 10000 MIPS generated task, the traffic is equal to 8 Gb/s. This traffic can be accommodated by the wired connection while the AP, in this case, consumes power to act as a coordinator.

The next section explains the results at different DRR values with high and low demands for one generated task. Both data sets are tested with single and distributed allocation strategies. A case was considered where all the processing nodes (CC, MF, LF, NF, and VN) are available, and the vehicular nodes exist with low density (CFVA-L).

## SCENARIO#1. One generated task with high processing demand

This section describes the results (for both single and distributed allocation strategies) when one generated task is under increasing demand from 1000 to 10000 MIPS.

First, Figures 28 and 29 show the total power consumption and processing allocation in each PN, respectively, with single allocation strategy.

Figure 28 shows that the total power consumption for single allocation strategy has relatively comparable behaviours for all DRR values. Moreover, Figure 29 shows that the processing allocation in all PNs remains the same for all DRR values between 0.8 and 0.08. This is due to the high data rate associated with these DRR values and with the high processing demands. As a result, no processing was allocated to the VEC despite the fact that a VN can process a single task up to 3200 MIPS. This is because the minimum processing demand for the single task (1000 MIPS) requires a minimum data rate of 80 Mb/s, which exceeds the connection capability of a VN.

Consequently, tasks were accommodated by NF and LF for all these DRR values (0.8 – 0.08).

With low DRR values, 0.04 and 0.02, the required data rate associated with low processing demands fit within the capacity of the VN connection. Therefore, tasks with 1000 MIPS, in DRR=0.04, and with 1000–3000 MIPS, in DRR=0.02, were allocated to the VEC, as shown in Figure 29. This explains the drop in power consumption for both ratios, saving up to 75% power compared to DRR=0.8, as depicted in Figure 28. However, the VEC allocation was terminated afterwards and the model followed the same allocation behaviour observed with other DRR values. This can be due to the VN connection limitation (in 0.04), or due to connection and processing capacity limitation (in 0.02). Yet, a saving of up to 42% is achieved, compared to DRR=0.8, as both ratios (0.02 and 0.04) result in a low data rate requirement and, therefore, less power consumption.

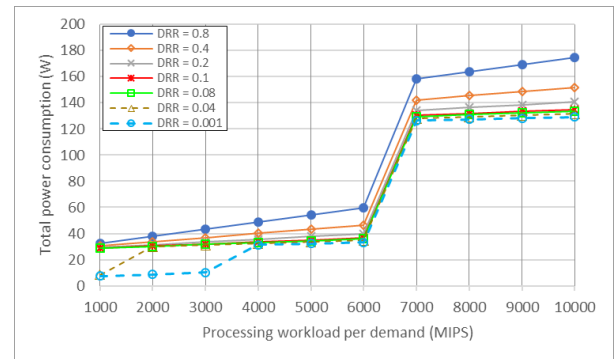

Fig. 28. Total power consumption in single allocation with high processing requirements

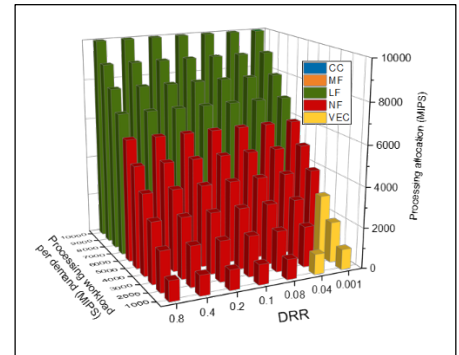

Fig. 29. Processing allocation in each PN in single allocation with high processing requirements

Figures 30 and 31 show the total power consumption and processing allocation in each PN, respectively, for the distributed allocation strategy. We observed that splitting the task can solve the VN connection limitation, as the required flow will be split proportionally with the given workload. Therefore, the optimization was able to achieve better utilization of the VEC with medium DRR values (0.2 – 0.08) and the low DRR values (0.04 and 0.02), as seen in Figure 31. For instance, with a DRR value of 0.2, the 200 Mb/s required data rate, accompanied by the 1000 MIPS task, was split among three VNs. Similarly, tasks with DRR equal to 0.1 (up to 2000

MIPS) and 0.08 (up to 3000 MIPS) were allocated to VEC by splitting the processing allocation among the available VNs. According to these results, and as shown by Figure 30, distributed allocation was able to achieve up to 63% power saving with medium DRR values, compared to single allocation. This saving increased to 63% – 68% in some cases in the three medium DRR values, as the optimum solution resulted in the use of the VEC with NF full utilization. This is a result of the higher power consumption which results from activating OLT and LF. The above matches the model behaviour with expanded architecture, as confirmed in Sections 1 and 2. With low DRR values (0.02 and 0.04), an improvement in VEC utilization and power savings is achieved. This is attributed to the low-demanding data rate per task in relation to the required processing workload and the low DRR values. The low data rate resulted in the optimization splitting the task among the available four VNs without exceeding the capacity of the VN connection. Using this allocation, the model saved power of 34% – 81% (with DRR 0.04), compared to single allocation. This saving percentage is based on the amount of extra workload allocated to VEC after utilizing the NF. With DRR=0.02, the power consumption saving increased further to up to 84%, as all tasks were allocated to the VEC.

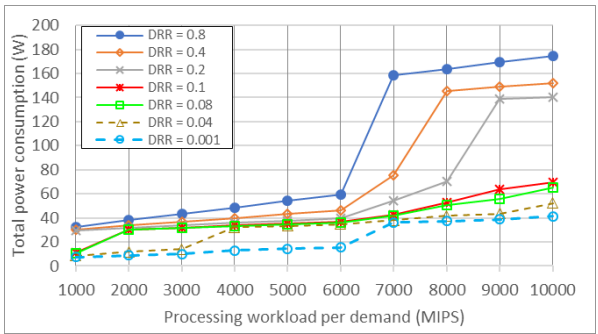

Fig. 30. Total power consumption in distributed allocation with high processing requirements

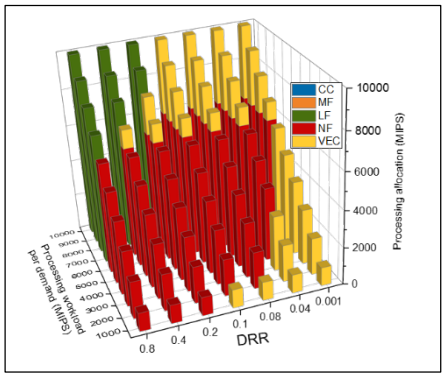

Fig. 31. Processing allocation in each PN in distributed allocation with high processing requirements

**SCENARIO#2. One generated task with low processing demand**

Similar to Scenario 1, this scenario evaluates a single generated task but with low processing demands, ranging from 100 to 1000 MIPS. As seen in Figures 32 and 33, the results showed similar processing allocation behaviors to Scenario 1.

However, as the low processing workload decreased the data rate requirement, a huge power saving is expected, and was confirmed, as seen in Figure 32. Figure 33 shows that DRR=0.8 caused a data rate bottleneck even with the lowest possible demand (i.e. 100 MIPS). However, with DRR=0.4, the VEC had a task allocation even with the lowest generated demand. In the same figure, we observe that the VEC utilization increased for medium DRR values (0.2–0.04), compared to Scenario 1. This is due to the low processing workload that can be satisfied by one VN processor. This VEC utilization stopped when the data rate exceeded the VN connection capacity, based on the DRR value. In both low DRR values (0.04 and 0.02), a full tasks allocation was achieved by the available VNs. This confirms that, regardless of the processing demand, VEC represents the most efficient PN, as long as its VNs and their connection can satisfy the tasks' requirements.

The results for the distributed allocation in Figures 34 and 35 show comparable improvement values in the power consumption and allocation to that achieved by Scenario 1 under high demands.

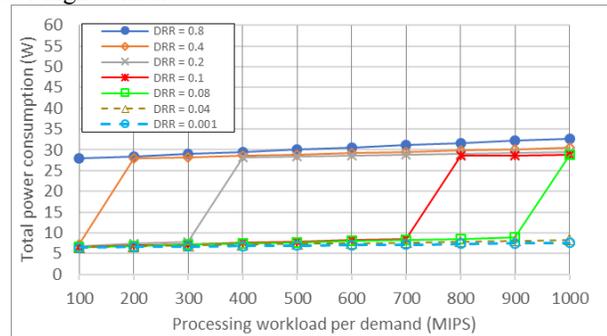

Fig. 32 Total power consumption in single allocation with low processing requirements

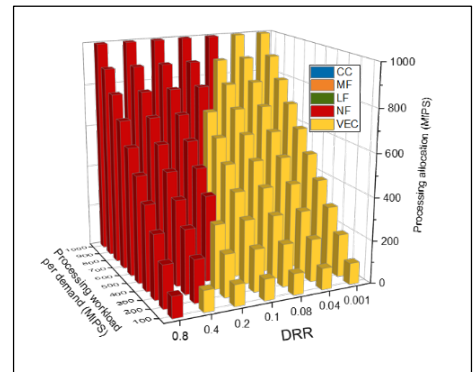

Fig. 33. Processing allocation in each PN in single allocation with low processing requirements

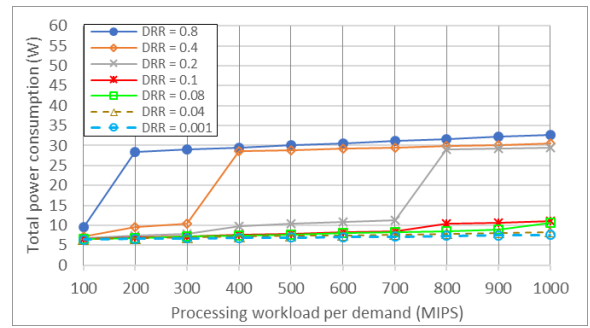

Fig. 34. Total power consumption in distributed allocation with low processing requirements

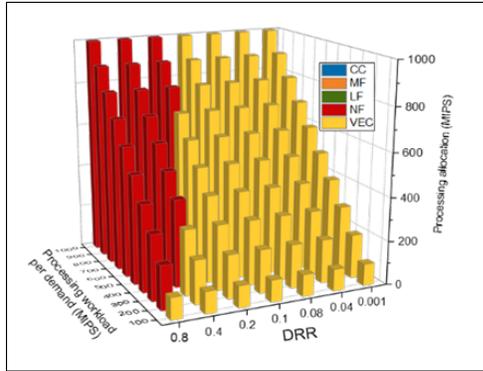

Fig. 35. Processing allocation in each PN in distributed allocation with low processing requirements

## VII. CONCLUSIONS

In this paper, we have investigated the processing allocation optimization problem in vehicular edge clouds integrated with central clouds and distributed fog processors. This architecture was evaluated through a MILP optimization model to minimize the total power consumption. The evaluation considered multiple cases to study the impact of workload volume, task generation density, vehicles density, and task allocation strategy (single and distributed). Two architectural designs were evaluated with single or multiple VEC clusters. The results of the investigation showed that vehicles with enough capacity turn out to be a very attractive option for processing the generated workload and saving power. As a result, a power savings up to 70% is achieved by allocating processing to the vehicles. This percentage varied based on the assessed scenario. Moreover, splitting the tasks between multiple vehicles achieved up to 71% power saving compared to the scenario with single allocation. The overhead power (idle power and PUE) of each processing server is a key factor that affects the allocation decisions. Accordingly, with high generated tasks associated with high volume demands, the central cloud becomes more efficient. It was also shown that expanding the access layer with multiple ONUs has minor effect on the allocation decisions, as the local VEC is always more efficient than the non-local VEC.

Future extensions will include developing a heuristic algorithm to approximate the constructed MILP models and hence allow the evaluation to be scaled up to increased number of processing nodes and vehicular nodes. Currently the nature of the MILP model allows the evaluation of small networks such as those considered in the scenarios presented, unless large computing resources are used in the MILP solution. Moreover, modelling the vehicles mobility and using prediction mechanisms rather than considering variable, but deterministic number of vehicles is of interest. This can mimic the stochastic behaviour of vehicles and can help evaluate the effect of dynamic arrival and departure of vehicles in the car parks and their impact on the reliability of the processing at vehicular nodes and the task completion success. This can incorporate task migration to other available vehicles or to fixed nodes.